\documentclass[a4paper,fleqn,usenatbib]{mnras}
\usepackage[T1]{fontenc}
\usepackage{ae,aecompl}

\usepackage{verbatim}
\usepackage{aas_macros}
\usepackage{hyperref}
\usepackage{natbib}
\usepackage{epsfig}
\usepackage[english]{babel}
\usepackage[normalem]{ulem}
\usepackage{xcolor}

\usepackage{graphicx}	
\usepackage{amsmath}	
\usepackage{amssymb}	

\newcommand\rsout{\bgroup\markoverwith{\textcolor{red}{\rule[0.5ex]{2pt}{1pt}}}\ULon}

\title[Black hole mass and galaxy properties in TNG]{The relationship between black hole mass and galaxy properties: Examining the black hole feedback model in IllustrisTNG}

\author[B. A. Terrazas et al.]{Bryan A. Terrazas$^{1}$\thanks{E-mail: bterraza@umich.edu},
Eric F. Bell$^{1}$,
Annalisa Pillepich$^{2}$,
Dylan Nelson$^{3}$, \newauthor
Rachel S. Somerville$^{4}$,
Shy Genel$^{4,5}$,
Rainer Weinberger$^{6}$,
M\'elanie Habouzit$^{4}$, \newauthor
Yuan Li$^{4,7}$,
Lars Hernquist$^{6}$, 
and Mark Vogelsberger$^{8}$
\vspace{2mm}
\\
$^{1}${Department of Astronomy, University of Michigan, Ann Arbor, MI 48109, USA}\\
$^{2}${Max-Planck-Institut f{\"u}r Astronomie, K{\"o}nigstuhl 17, 69117 Heidelberg, Germany}\\
$^{3}${Max-Planck-Institut f{\"u}r Astrophysik, Karl-Schwarzschild-Str. 1, D-85748, Garching, Germany}\\
$^{4}${Center for Computational Astrophysics, Flatiron Institute, 162 Fifth Ave., New York, NY 10027, USA}\\
$^{5}${Columbia Astrophysics Laboratory, Columbia University, 550 West 120th Street, New York, NY 10027, USA}\\
$^{6}${Institute for Theory and Computation, Harvard-Smithsonian Center for Astrophysics, 60 Garden Street, Cambridge, MA 02138, USA}\\
$^{7}${Astronomy Department, University of California, Berkeley, CA 94720, USA}\\
$^{8}${Department of Physics, Massachusetts Institute of Technology, Cambridge, MA 02139, USA}
}

\pubyear{2019}

\begin{document}
\label{firstpage}
\pagerange{\pageref{firstpage}--\pageref{lastpage}}
\maketitle

\begin{abstract}
Supermassive black hole feedback is thought to be responsible for the lack of star formation, or quiescence, in a significant fraction of galaxies. We explore how observable correlations between the specific star formation rate (sSFR), stellar mass (M$_{\rm{star}}$), and black hole mass (M$_{\rm{BH}}$) are sensitive to the physics of black hole feedback in a galaxy formation model. We use the IllustrisTNG simulation suite, specifically the TNG100 simulation and ten model variations that alter the parameters of the black hole model. Focusing on central galaxies at $z = 0$ with M$_{\rm{star}} > 10^{10}$ M$_{\odot}$, we find that the sSFR of galaxies in IllustrisTNG decreases once the energy from black hole kinetic winds at low accretion rates becomes larger than the gravitational binding energy of gas within the galaxy stellar radius. This occurs at a particular M$_{\rm{BH}}$ threshold above which galaxies are found to sharply transition from being mostly star-forming to mostly quiescent. As a result of this behavior, the fraction of quiescent galaxies as a function of M$_{\rm{star}}$ is sensitive to both the normalization of the M$_{\rm{BH}}$-M$_{\rm{star}}$ relation and the M$_{\rm{BH}}$ threshold for quiescence in IllustrisTNG. Finally, we compare these model results to observations of 91 central galaxies with dynamical M$_{\rm{BH}}$ measurements with the caveat that this sample is not representative of the whole galaxy population. While IllustrisTNG reproduces the observed trend that quiescent galaxies host more massive black holes, the observations exhibit a broader scatter in M$_{\rm{BH}}$ at a given M$_{\rm{star}}$ and show a smoother decline in sSFR with M$_{\rm{BH}}$. 
\end{abstract}



\section{Introduction}
\label{sec:intro}

In the last 10 billion years, the amount of new star formation in the Universe has decreased substantially. Observational surveys of galaxy populations at different epochs have also shown a gradual increase in the number of galaxies whose light is dominated by older stellar populations signaling a lack of new stars in these systems \citep[e.g.,][]{bwm2004, bgb2004, wqf2009, mms2013, iml2013, tqt2014}. Studies of quiescence, i.e. the state of reduced star formation activity, in galaxies have aimed at pinpointing a physical mechanism responsible for this behavior. Many of these studies, including this work, focus on \textit{central} galaxies situated at the centres of their dark matter haloes, since satellites undergo additional environmental processes that may affect their star formation activity. Yet, clear identification of the mechanisms behind quiescence in these central galaxies has proved elusive, as galaxy evolution is complex and many physical processes act in concert to shape the properties of galaxies. 


One observational approach to understanding quiescence in central galaxies focuses on measuring correlations between galaxy properties and quiescence. For example, galaxies with high stellar masses \citep[e.g.,][]{khw2003}, bulge-dominated morphologies \citep[e.g.,][]{bvp2012, bme2014}, high halo masses \citep[e.g.,][]{w2018}, high S\'ersic indices \citep[e.g.,][]{cfk2012}, high central stellar surface densities \citep[e.g.,][]{wdf2015}, and more massive black holes \citep[e.g.,][]{tbh2016b, tbw2017, mbr2018} have a higher likelihood of being quiescent. Several mechanisms attempting to explain these correlations have been proposed, such as supernovae/stellar winds \citep[e.g.,][]{tw2005, cvk2015}, gravitational heating \citep[e.g.,][]{jno2009}, morphological quenching \citep[e.g.,][]{mbt2009}, halo mass quenching \citep[e.g.,][]{db2006, cdd2006, bdn2007}, and central black hole feedback \citep[e.g.,][]{kh2000, dsh2005, csw2006, bbm2006, shc2008, f2012}.



In a cosmological context, gas that falls into a dark matter halo is able to lose energy and cool via dissipative processes to form a disc of cold gas at the bottom of the halo's potential well \citep[e.g.][]{s1977, wr1978, fe1980, kg1991}. The gas that accumulates in the disc fragments, cools, and condenses into molecular clouds which eventually collapse into new stars. 

One natural feedback channel that limits star formation is stellar feedback from star formation itself. Radiation from stars, stellar winds, and supernovae regulate the production of new stars by limiting the amount of cold dense gas \citep[e.g.,][]{s2003, sh2003, hqm2011}. This form of feedback, however, is insufficient for producing quiescence in massive galaxies. Recent models show that the gas expelled from galaxies by stellar feedback is reincorporated to provide fuel for continued star formation \citep[e.g.,][]{dpp2016, ptr2017, shh2018, cso2018}, a behavior that is more efficient for more massive systems \citep{odk2010, bbc2012, cdg2016, mkf2015}. Thus, a picture emerges of galaxies undergoing a cycle of ejection and reincorporation that regulates cooling and star formation via stellar feedback similar to various `bathtub' models that have been proposed \citep{bdg2010, dvd2010, dfo2012, pm2014, dm2014, bla2014}. 

In this physically-motivated framework, quiescence can be defined as the consequence resulting from the disruption of this cycle. This disruption can take the form of `ejective' and/or `preventative' feedback (see Section 3.3 of \citealt{sd2015} for a full review). Ejective feedback pushes gas out of galaxies which may otherwise continue forming stars. Preventative feedback prevents galaxies from accumulating star-forming gas from the cooling of the circumgalactic medium. While both processes likely play a role in suppressing star formation, any proposed mechanism for long-term quiescence must be at least partly preventative in nature since the accretion of new gas would re-establish star formation in the galaxy.


A popular mechanism for producing quiescence in recent physics-based galaxy formation models is black hole feedback \citep{bbm2006, csw2006, ssd2007, shc2008, gwl2010, bs2011, vgs2014, hwt2015, sd2015, scb2015, dpp2016, msb2017, bsf2017, wsp2018}. Theoretically, accretion onto a black hole has the potential to release an enormous amount of energy into the surrounding medium. Observationally, activity from central supermassive black holes has been seen in many forms. 

When black holes undergo episodes of high accretion rate feedback, the radiation pressure created by the luminous accretion disc is thought to result in large-scale outflows. Many X-ray luminous active galactic nuclei (AGN) are observed to host ionized gas outflows in support of this idea \citep{hmv1981, csk2010, vhd2011, cms2014, wbs2016, rgv2017}. At lower accretion rates, black holes are thought to produce jets from their accretion discs which propel low-density buoyant bubbles into the atmospheres of galaxies. Indeed, large extended radio-emitting lobes that create cavities within the surrounding hot gas visible in X-ray maps have been observed around massive galaxies in various environments from small groups to large clusters \citep[e.g.,][]{brm2004, mn2007, f2012, swm2016, wmc2019}. Some galaxies have been observed to exhibit signs of both a bright X-ray point source in conjunction with extended radio emission, further complicating the issue of the accretion physics that could produce these varied effects on the gas within and around galaxies \citep{kvx2006, yzk2008, bfc2015, cat2017}. More recently, some galaxies with low luminosity AGN have been observed to produce bisymmetric winds of ionized gas which are thought to affect star formation in the galaxy \citep{cbc2016, pms2018}. Finally, our own Milky Way galaxy also hosts evidence of a possible relic from black hole feedback in the form of Fermi bubbles \citep{ssf2010, gm2012}. 


The black hole accretion physics occurring at parsec-scales and producing these observed forms of black hole activity is not well understood. As a result, there is substantial freedom in the subgrid physics used between different cosmological simulations to model the effects black holes may have on the surrounding gas. This can result in important differences in the star formation histories, gas content, and stellar mass distributions of simulated galaxy populations which can then be compared to observational data \citep[e.g.,][]{tbh2016b, tbw2017, bme2016, bbt2019}. As such, understanding the details of how small-scale subgrid physics affects large-scale galaxy population statistics in models can improve our understanding of the feedback mechanisms that may produce quiescence in the real Universe. 

With these considerations in mind, the central challenge this work seeks to address is understanding how observed galaxy correlations between specific star formation rate (sSFR), stellar mass (M$_{\rm{star}}$), and black hole mass (M$_{\rm{BH}}$) at $z = 0$ can be interpreted in light of the results from a physical model of black hole feedback. In this work, we use the IllustrisTNG simulation suite \citep[TNG; ][]{spp2018, mvp2018, nps2018, pnh2018, dnps2018} in order to explore this question, focusing on central galaxies with M$_{\rm{star}} > 10^{10}$ M$_{\odot}$. TNG uses black hole feedback in order to suppress the SFRs and M$_{\rm{star}}$ content of galaxies with M$_{\rm{star}} \gtrsim 10^{10}$ M$_{\odot}$ \citep{wsh2017, wsp2018}. Thus, as we will show, the sSFR, M$_{\rm{star}}$, and M$_{\rm{BH}}$ of galaxies are causally related to one another through the model's mechanism for quiescence.

One important feature of TNG for our purposes is the availability of dozens of TNG model variations, first introduced in \citet{psn2018}. We use these model variations to study the impact of different parameter choices, as is similarly done in \citet{wsh2017}. By using the model variations explicitly related to black hole feedback, we can explore how changes to the model can affect TNG's galaxy population statistics. We aim to link these differences in observable correlations to the effects of black hole feedback on the physical properties of the gas within the galaxy and in the surrounding circumgalactic medium. Thus, our work will show how these observed correlations may be physically interpreted in the real Universe.

We organize our results as follows: Section~\ref{sec:tng} describes the TNG simulation suite, the model variations, the black hole physics model, and our definitions of galaxy properties in the model. Section~\ref{sec:quiescence} describes the necessary conditions for quiescence in TNG using three of the model variations. Section~\ref{sec:physics} describes the effects of black hole feedback on gas and outlines a phenomenological framework for the physics of quiescence in TNG. Section~\ref{sec:compobs} compares the model results to observational correlations, illuminating how the sSFR, M$_{\rm{star}}$, and M$_{\rm{BH}}$ can encode information on the physics behind quiescence within the context of black hole feedback. Section~\ref{sec:implications} uses model variations to illuminate how observables may be affected by changes in the way black hole feedback operates. In Section~\ref{sec:disc} we reflect on our results and present future outlooks. Section~\ref{sec:conclusions} summarizes our findings and contains our concluding remarks.


\begin{table*}
\centering
\caption{The list of model variations we use for this work. Columns: (1) model number, (2) model name, (3) parameter changed, (4) fiducial model value, (5) changed value, (6) short description of the alteration. See Section~\ref{sec:bhs} for more information on each of the model parameters in column (3).}
\label{tab:models}
\begin{tabular}{llllll}
\hline Model & Name (2) & Parameter & Fiducial & Changed & Description (6)  \\ 
No. (1) &  & changed (3) & value (4) & value (5) & \\ \hline
0000 & FiducialModel & n/a & n/a & n/a & Same physics as the larger TNG volumes \\
2201 & NoBHs & n/a & n/a & n/a & Black holes turned off \\
3000 & NoBHwinds & $\chi_{0}$ & 0.002 & 0 & Thermal mode at all $\dot{\rm{M}}_{\rm{BH}}$ \\
3101 & LowKinEff & $\varepsilon_{\rm{f,kin,max}}$ & 0.2 & 0.05 & Lower gas coupling efficiency for BH kinetic feedback \\
3102 & HighKinEff & $\varepsilon_{\rm{f,kin,max}}$ & 0.2 & 0.8 & Higher gas coupling efficiency for BH kinetic feedback \\
3103 & HighThermEff & $\varepsilon_{\rm{f,high}}$ & 0.1 & 0.4 & Higher gas coupling efficiency for BH thermal feedback \\
3104 & LowThermEff & $\varepsilon_{\rm{f,high}}$ & 0.1 & 0.05 & Lower gas coupling efficiency for BH thermal feedback \\
3301 & HighChi0 & $\chi_{0}$ & 0.002 & 0.008 & Higher Eddington ratio threshold normalization \\
3302 & LowChi0 & $\chi_{0}$ & 0.002 & 0.0005 & Lower Eddington ratio threshold normalization \\
3403 & HighMpiv & $M_{\rm{piv}}$ & 1e8 $\rm{M}_{\odot}$ & 4e8 $\rm{M}_{\odot}$ & Higher pivot mass in Eddington ratio threshold \\
3404 & LowMpiv & $M_{\rm{piv}}$ & 1e8 $\rm{M}_{\odot}$ & 2.5e7 $\rm{M}_{\odot}$ & Lower pivot mass in Eddington ratio threshold \\ \hline
\end{tabular}
\end{table*}

\section{The Illustris TNG Simulation Suite}
\label{sec:tng}

The IllustrisTNG project is a large-scale cosmological and gravo-magneto-hydrodynamical simulation suite of galaxy formation in a $\Lambda$CDM Universe \citep{spp2018, mvp2018, nps2018, pnh2018, dnps2018}. TNG is the descendant of the original Illustris project \citep{vgs2014, vgs2014b, gvs2014, svg2015}, modifying and adding numerous features with the goal to improve the agreement between the simulation and observational results by providing a comprehensive physical model of galaxy formation from the early Universe to the present day. The simulation uses the $\textsc{Arepo}$ code to solve the equations of ideal magnetohydrodynamics and self-gravity on a moving, unstructured mesh \citep{s2010, pbs2011, psb2016}. For more information on the numerical aspects of the TNG model, we refer the reader to \citet{psn2018} and \citet{wsh2017}.

The simulation suite contains three simulation volumes TNG50, TNG100, and TNG300 sized at 51.7$^{3}$, 110.7$^{3}$, and 302.6$^{3}$ comoving Mpc$^{3}$ volumes, respectively. In this work, we will be focusing on results from TNG100 as it has roughly the same resolution (within a factor of 2) as the model variations which we will use extensively in this paper and which are detailed in Section~\ref{sec:modelvariations}. This will allow a comparison without the complications of resolution effects present in TNG300 (see Appendix~\ref{sec:tng300}). TNG100 has $2\times1820^{3}$ initial resolution elements with a baryonic mass resolution of $1.4\times10^{6}$ M$_{\odot}$ and gravitational softening length of 0.74 kpc at $z = 0$. The cosmological parameters of the model are based on \citet{planck2016} with a matter density $\Omega_{M,0} = 0.3089$, baryon density $\Omega_{b,0} = 0.0486$, dark energy density $\Omega_{\Lambda,0} =  0.6911$, Hubble constant H$_{0}$ = 67.74 km/s/Mpc, power spectrum normalization factor $\sigma_{8} = 0.8159$, and spectral index $n_{s} = 0.9667$.

TNG models the physics of primordial and metal-line gas cooling, magnetic fields, star formation, stellar evolution and feedback, chemical enrichment, and black hole growth and feedback. There are several key differences between the original Illustris model and TNG which are described in \citet{psn2018}. For our purposes, the most relevant modification is how black hole feedback operates at low accretion rates. Instead of the thermal bubble model implemented by the original Illustris model and described in \citet{ssd2007, svg2015}, TNG adopts a kinetic wind model that inputs kinetic energy originating at the black hole into nearby gas particles \citep{wsh2017}. The primary motivation for changing the physical prescription for black hole feedback was to prevent the ejection of large amounts of gas from the haloes of intermediate to high mass galaxies. The amount of halo gas around many of these galaxies in Illustris is much lower than observations suggest \citep[see Figure 10 in][]{gvs2014}, yet the mass and state of the interstellar medium gas is such that a clear colour bimodality as compared to SDSS data is not present (see Figure 14 in \citealt{vgs2014}). This led to the implementation of a modified form of feedback.

\subsection{The formation, growth, and feedback of black holes in TNG}
\label{sec:bhs}

Here we provide a brief description of the black hole model in TNG. For full details, see \citet{wsh2017}.

Black holes are placed at the centre of a halo's potential well with a seed mass of M$_{\rm{seed}} = 1.18 \times 10^{6}$ M$_{\odot}$ once a halo grows past a threshold mass of M$_{\rm{h}} = 7.38 \times 10^{10}$ M$_{\odot}$. Once seeded, black holes grow either through accretion at the Eddington-limited Bondi accretion rate or by merging with other black holes during a galaxy merger. Additionally, black holes are made to stay at the potential minimum of their host subhaloes at each global integration time step in order to avoid numerical effects that may displace them.



TNG employs a M$_{\rm{BH}}$-dependent Eddington ratio threshold for determining whether a black hole provides either pure thermal or pure kinetic mode feedback energy to the galaxy. Kinetic mode feedback in TNG is turned on when a galaxy's black hole accretion rate drops below an Eddington ratio of
\begin{equation}
\begin{aligned}
\chi = \rm{min}\bigg[ \chi_{0} \Big(\frac{\textit{M}_{\rm{BH}}}{\textit{M}_{\rm{\rm{piv}}}}\Big)^{\beta}, \chi_{\rm{max}}\bigg],
\label{eqn:chi}
\end{aligned}
\end{equation}
where $M_{\rm{BH}}$ is the black hole mass and the parameters for the fiducial model are $\chi_{0} = 0.002$, $M_{\rm{piv}} = 10^{8}$ M$_{\odot}$, $\beta = 2$, and $\chi_{\rm{max}} = 0.1$ (refer to Equation 5 and Figure 6 of \citet{wsh2017}). 

Apart from $\chi_{\rm{max}}$, which is chosen to be the canonical, observationally-suggested value of 0.1 for the Eddington ratio of black holes in quasars \citep{yt2002}, none of the other parameter values were adopted based on empirical evidence or other theoretical models of black hole physics and accretion. The parameters have been chosen to ensure that the TNG model returns the observed stellar mass content of massive haloes and the observed location of the knee in the stellar mass function at the current epoch.

Thermal mode feedback energy is parameterized as $\dot{E}_{\rm{thermal}} = \varepsilon_{\rm{f,high}} \varepsilon_{\rm{r}} \dot{\textit{M}}_{\rm{BH}} \textit{c}^{2}$, where $\varepsilon_{\rm{f,high}}$ is the fraction of thermal energy that couples to the surrounding gas (set to 0.1 for the fiducial model) and $\varepsilon_{\rm{r}}$ is the black hole radiative efficiency (set to 0.2 for the fiducial model). Kinetic mode feedback energy is parameterized as $\dot{E}_{\rm{kinetic}} = \varepsilon_{\rm{f,kinetic}} \dot{\textit{M}}_{\rm{BH}} \textit{c}^{2}$, where $\varepsilon_{\rm{f,kinetic}}$ is the fraction of kinetic energy that couples to the surrounding gas (set to a maximum value of $\varepsilon_{\rm{f,kin,max}} = $ 0.2 for the fiducial model, see Equation 9 of \citealt{wsh2017} for a full description). Once the black hole accumulates enough energy in this mode to reach a minimum energy threshold, $E_{\rm{inj,min}}$, the gas immediately surrounding the black hole receives a momentum kick in a random direction away from the black hole. 

\subsection{Model Variations}
\label{sec:modelvariations}

Along with TNG100, there are dozens of smaller simulations sized at 36.9$^{3}$ comoving Mpc$^{3}$ volumes that vary individual parameters of the model \citep{psn2018}. These simulations have $2\times512^{3}$ initial resolution elements with roughly the same baryonic mass resolution ($2.4\times10^{6}$ M$_{\odot}$) and gravitational softening length (0.74 kpc at $z = 0$) as TNG100. Each of these simulations have identical cosmological initial conditions resulting in similar dark matter structures and galaxy placement throughout the volume. This allows a galaxy-by-galaxy comparison of the model variations, providing us with a powerful tool to explore the effects of each component of the model relevant for quiescence.

In Table~\ref{tab:models} we describe the model variations that we use in this work. The first row describes a model with the same physics as TNG100 but with the same initial conditions, resolution, and volume as all other model variations. We will refer to this run as the FiducialModel simulation. Each of the other model variations alter the black hole feedback model described in Section~\ref{sec:bhs}. These variations will affect how galaxies populate the M$_{\rm{BH}}$-M$_{\rm{star}}$-sSFR parameter space, as we will show in later sections. The details of each altered parameter in the third column are given in Section~\ref{sec:bhs}, with a brief description of the change in the sixth column. We will refer to each simulation according to their names in the second column of Table~\ref{tab:models}.

\subsection{Definitions of physical properties in TNG}
\label{sec:definitions}

We calculate the M$_{\rm{star}}$ and SFR within the galaxy radius, defined as twice the stellar half mass radius. SFRs are calculated by averaging the star formation activity in the last 200 Myr in order to reasonably compare our results to observational SFR indicators. Our results remain qualitatively the same whether we use instantaneous SFRs or those averaged over 50, 100, or 200 Myr.

We define star-forming and quiescent galaxies to be those with sSFRs above or below $10^{-11}$ yr$^{-1}$, respectively. This provides a consistent separation between galaxies on and off the star forming main sequence for both observations and simulations at $z = 0$.

TNG100 and the model variations are sensitive to 200 Myr-averaged SFRs above $\sim10^{-2.5}$ M$_{\odot}$ yr$^{-1}$, below which star formation is unresolved \citep[see Figure 2 in][]{dpn2019}. This value is related to the mass resolution limit of a single stellar particle in TNG100 and the model variations. Since we focus on galaxies with M$_{\rm{star}} > 10^{10}$ M$_{\odot}$, any values for sSFR $< 10^{-12.5}$ yr$^{-1}$ will be taken as an upper limit. This ensures that all sSFRs above this limit are resolved. Upper limits will be shown as a value chosen from a gaussian distribution centred at sSFR $= 10^{-12.5}$ yr$^{-1}$ in order to visualize what would be observed as non-detections.

Black hole parameters are taken for the most massive black hole at the centre of the galaxy. Halo properties such as gas cooling rates ($\mathcal{C}_{\rm{halo}}$), gas masses (M$_{\rm{gas,halo}}$), and dark matter halo masses (M$_{\rm{DM,halo}}$) are the sum values for each cell within the radius at which the density within is 200 times the critical density of the Universe, $R_{200\rm{c}}$. Halo gas cooling times (t$_{\rm{cool}}$) are the average of gas cells within $R_{200\rm{c}}$.

Additionally, our analysis will use a measure of the gravitational binding energy of gas within the galaxy, which we define as
\begin{equation}
\begin{aligned}
E_{\rm{bind, gal}}(<r_{\rm{gal}}) = \frac{1}{2} \sum\limits_{\rm{g}(<r_{\rm{gal}})} m_{\rm{g}} \phi_{\rm{g}},
\end{aligned}
\label{eqn:be}
\end{equation}
where $r_{\rm{gal}}$ is twice the stellar half mass radius, g indexes gas cells in the simulation, $m_{\rm{g}}$ is the mass of the gas cell, and $\phi_{\rm{g}}$ is the gravitational potential felt at the gas cell's position within the galaxy. This value represents the energy needed to unbind those gas cells from the galaxy's position to infinity, taking into account not only the mass within the galaxy but also the mass within the halo. We choose to calculate the binding energy of gas within this radius in order to match the radius within which the SFR and M$_{\rm{star}}$ are calculated.

\section{The necessary conditions for quiescence}
\label{sec:quiescence}

\begin{figure}
\centering 
\epsfxsize=8.1cm
\epsfbox{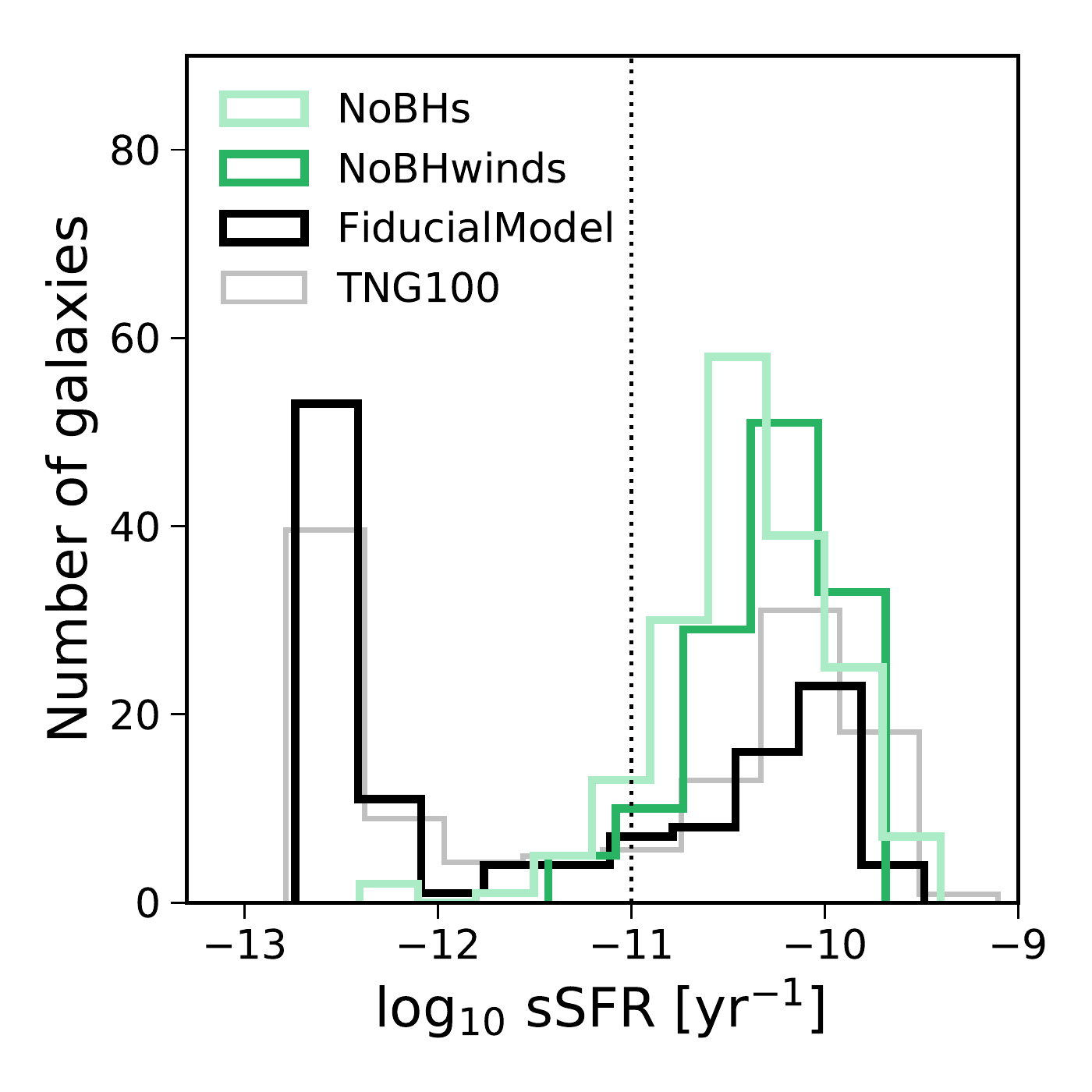}  
\caption{Histograms of sSFR for central galaxies with M$_{\rm{star}} > 10^{10}$ M$_{\odot}$ at $z = 0$ for the model variations with no black holes (light green), thermal mode at all accretion rates (dark green), and both thermal and kinetic modes included as in the fiducial model (black). The number of galaxies shown in each of these histograms is roughly the same. Due to the smaller box size, there are very few galaxies with M$_{\rm{DM,halo}} > 10^{13}$ M$_{\odot}$. TNG100 scaled to the volume of the smaller box simulations is shown in gray. The star forming main sequence is present in all simulations but a quiescent population only appears in the models with kinetic mode black hole feedback.}
\label{fig:ssfrhist}
\end{figure}

We begin by exploring the necessary conditions under which quiescence can occur for central galaxies with M$_{\rm{star}} > 10^{10}$ M$_{\odot}$ in the context of the TNG model. In Figure~\ref{fig:ssfrhist}, we show histograms of the number of galaxies as a function of sSFR for three model variations at $z = 0$. Due to their smaller box size, there are very few galaxies with M$_{\rm{DM,halo}} > 10^{13}$ M$_{\odot}$. We also plot galaxies from TNG100 (renormalized to match the box size of the FiducialModel run) in gray to show the results from a larger sample size with more massive haloes.

In light green, we show the TNG model variation with no black holes, where the only feedback channel is stellar feedback. This simulation results in a distribution of galaxies centred at sSFR $\sim 10^{-10}$ yr$^{-1}$, denoting the existence of the star forming main sequence.

In dark green, we show the model variation with black holes included where thermal mode black hole feedback operates for all accretion rates (the NoBHwinds model in Table~\ref{tab:models}). In TNG, a black hole accreting in the thermal mode injects pure thermal energy into the surrounding gas particles. \citet{wsp2018} show that the total amount of thermal energy released in this mode can be large yet has little effect on cooling and the galaxy's SFR. Thermally injected energy, as implemented for black hole thermal mode feedback in TNG, is more likely to be immediately radiated away rather than have any lasting impact on the thermodynamic properties of the gas, especially for dense gas where the cooling times are short \citep{nw1993, kwh1996}. This is likely due, at least in part, to a numerical effect resulting from the limited resolution of the simulation and leading to the `over-cooling' problem \citep[e.g.,][]{sh2002}. The pileup of galaxies at sSFR $\sim10^{-10}$ yr$^{-1}$ illustrates the inability of thermal energy injection, at least in the continuous fashion with which it is implemented within the TNG model, to prevent gas cooling and star formation in TNG galaxies.

\begin{figure}
\centering
\epsfxsize=8.4cm
\epsfbox{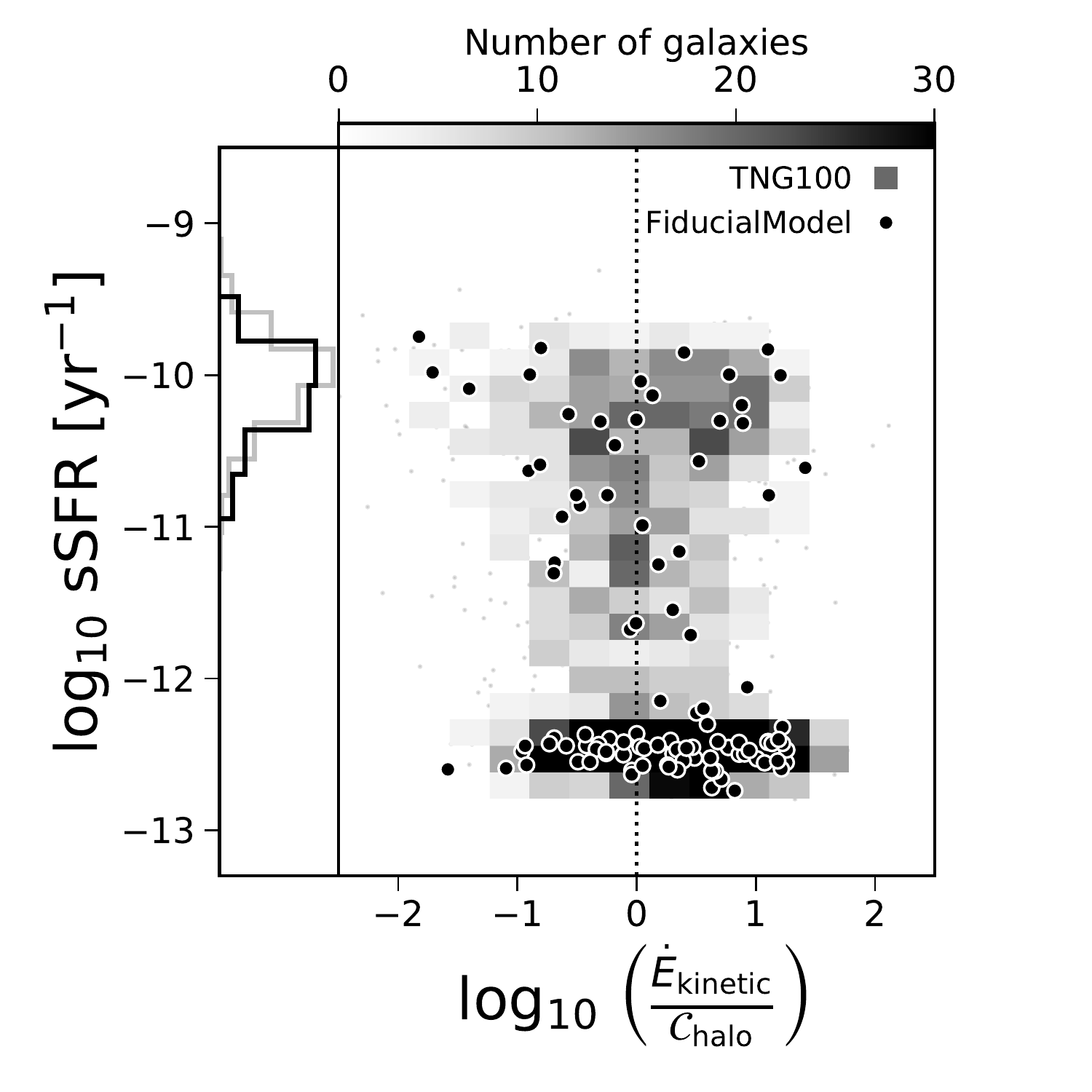}
\caption{sSFR as a function of the ratio between the instantaneous black hole wind energy injection rate, $\dot{E}_{\rm{kinetic}}$, and the instantaneous halo gas cooling rate, $\mathcal{C}_{\rm{halo}}$. The vertical dotted line shows where these two energy rates equal. The grayscale heatmap shows the distribution of galaxies in TNG100 and the black points show galaxies in the FiducialModel simulation at $z = 0$. The histogram shows the distribution of sSFR for galaxies with no black hole wind energy injection, $\dot{E}_{\rm{kinetic}} = 0$, for the TNG100 (gray, scaled to the volume of the FiducialModel run) and FiducialModel (black) simulations. While kinetic winds are required for quiescence, comparing the kinetic energy injection rate to the halo gas cooling rate is a poor indicator of a galaxy's star formation properties.}
\label{fig:energyrates}
\end{figure}

The black histogram shows galaxies in the FiducialModel simulation, where kinetic mode black hole feedback is turned on at low accretion rates. This black histogram along with the TNG100 results in gray show a significant population of quiescent galaxies at low sSFRs, most of which have upper limit values placed at sSFR $\sim 10^{-12.5}$ yr$^{-1}$. These results show that the existence of a quiescent population in TNG depends on the model's implementation of low accretion rate kinetic mode feedback from the central black hole. Similar conclusions have been made in previous studies of TNG both in terms of colour \citep{wsh2017, dnps2018} and sSFR \citep{wsp2018}.

%
\begin{figure*}
\centering
\epsfxsize=13.5cm
\epsfbox{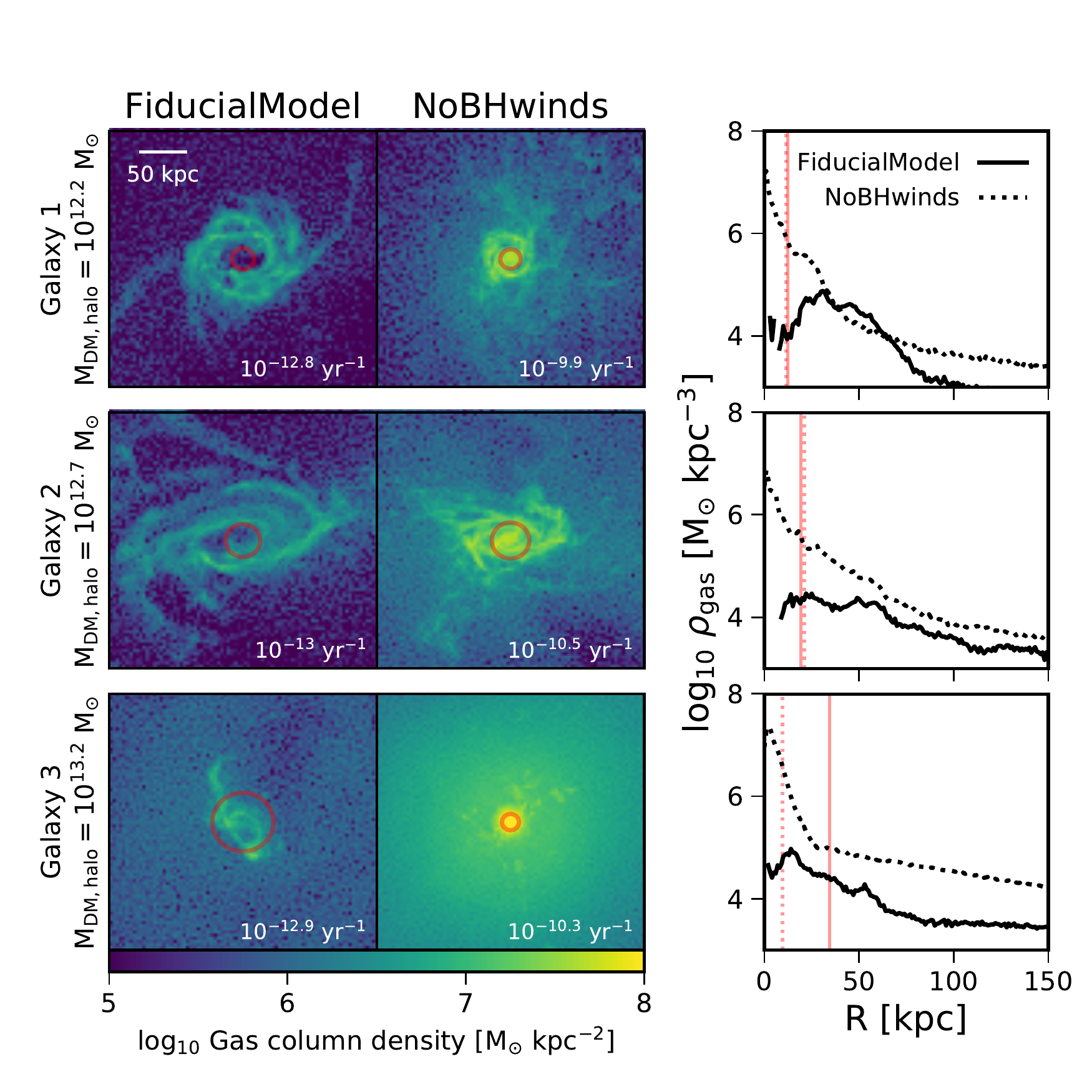}
\caption{\textit{Left panels}: The leftmost galaxy images show the projected (across 300 kpc) gas column density distributions of three quiescent galaxies in the FiducialModel simulation at $z = 0$ ordered by increasing mass from top to bottom. The images to the right show galaxies identified to be the centrals of the same haloes as those on the left but in the NoBHwinds simulation. The values on the lower right indicate the sSFR of each galaxy. Each image is 300$\times$300 kpc in size in order to directly compare the distribution of gas. The red circles indicate the galaxy radius, defined as twice the stellar half mass radius. \textit{Right panels:} The radial gas density distributions of the galaxies in the FiducialModel (solid lines) and NoBHwinds (dotted lines) simulations. The vertical red lines indicate the galaxy radius. The density of gas in the central regions of galaxies experiencing kinetic winds is depleted by orders of magnitude.}
\label{fig:gasdistcomp}
\end{figure*}

\section{The physics of quiescence from black hole-driven kinetic winds}
\label{sec:physics}

In this section we describe how black hole-driven kinetic winds employed by TNG affect the gas within and around galaxies in order to produce quiescence.

\subsection{Comparing TNG with semi-analytic approaches to quiescence}
\label{sec:sam}

Semi-analytic models of galaxy formation often use prescriptions of gas cooling and heating rates in order to determine the net rate of cold gas accreted onto a galaxy \citep[e.g.,][]{dlb2007, shc2008, gwl2010, hwt2015}. In these models, once the heating rate from black hole feedback equals or exceeds the halo gas cooling rate, cold gas accretion onto the galaxy halts and star formation subsequently shuts off. This describes a method to model \textit{preventative} black hole feedback, where fuel for star formation is prevented from reaching the galaxy.

In order to test whether a similar physical scenario occurs for TNG galaxies, we show the sSFR as a function of the ratio between the instantaneous energy released from black hole-driven kinetic feedback, $\dot{E}_{\rm{kinetic}}$ (see Section~\ref{sec:bhs}), and the instantaneous cooling rate of the gas halo, $\mathcal{C}_{\rm{halo}}$\footnote{Due to the lack of numerical resolution, star forming cells are described by an effective equation of state model once a density threshold is reached \citep{sh2003}. As such, cooling rates for these cells are excluded in the calculation of the halo cooling rate. Star forming cells account for a negligible amount of mass in the gas halo.}, at $z = 0$. The distribution of TNG100 galaxies is shown as a grayscale heatmap and FiducialModel galaxies are shown as black points. The vertical dotted line indicates where the two rates are equal. Since many galaxies are not in the kinetic mode (with $\dot{E}_{\rm{kinetic}} = 0$) and therefore cannot be represented on this plot, we show the distribution of their sSFRs as a histogram to the left of the main plot for both the TNG100 (gray, scaled to the volume of the FiducialModel run) and FiducialModel (black) simulations.

We find that all galaxies that are not currently experiencing kinetic mode black hole feedback (with $\dot{E}_{\rm{kinetic}} = 0$) are actively forming stars (left panel histogram in Figure~\ref{fig:energyrates}). The main panel of this figure shows that galaxies must be experiencing kinetic mode black hole feedback in order to be quiescent, in agreement with our results from Section~\ref{sec:quiescence}. 

However, a significant number of galaxies releasing black hole-driven kinetic winds are star-forming. In fact, whether these galaxies are star-forming or quiescent does not correlate tightly with whether the black hole wind energy greatly exceeds or falls below the cooling rate of the gas halo. As such, the TNG simulation cannot be easily described using the logic employed by semi-analytic models. We note that the comparison between these `heating' and `cooling' rates in TNG differ in detail to those analytically determined in semi-analytic approaches. This is due to the fact that these rates in TNG are sensitive to internal gas hydrodynamics whereas in semi-analytic models these rates are calculated using global galaxy and halo properties. Even so, this analysis indicates differences between how black hole feedback energy is transferred to the surrounding gas to produce quiescence in TNG and how it is transferred in purely preventative semi-analytic black hole feedback models.

\subsection{Gas distributions in TNG galaxies}
\label{sec:casestudy}

A useful approach for understanding the effects of black hole feedback on the gas within and around galaxies in TNG is to visualize and quantify the distribution of this gas. We do this using the model variations described in Section~\ref{sec:modelvariations}. Figure~\ref{fig:gasdistcomp} shows the gas column densities of three quiescent galaxies of increasing mass at $z = 0$ in the FiducialModel simulation (left images) and their direct counterparts in the NoBHwinds simulation (right images). Their radial density profiles are shown in the rightmost panels. We match these galaxies between the two model variations by the position of its dark matter halo and by ensuring that at least half of the dark matter particles have the same IDs. Each panel is at the same spatial scale of 300$\times$300 kpc. The circle at the centre of each figure depicts twice the stellar half mass radius of the galaxy from its centre in order to show where a galaxy's visible matter would lie in the image.

We find that there is much less dense gas in the central regions of the galaxies undergoing kinetic feedback. The discs of these galaxies are extended and disturbed, showing that black hole kinetic winds produce outflows that push gas out of galaxies, in agreement with recent TNG50 results described in \citet{nps2019}. Additionally, the density of the gas extending past the disc and into the gas halo is also depleted. We find that the same galaxies in the NoBHwinds simulation are star-forming, have retained dense gas within the galaxy's radius, and have centrally peaked radial density profiles (dotted lines in the right hand panels).

\begin{figure}
\centering
\epsfxsize=7.5cm
\epsfbox{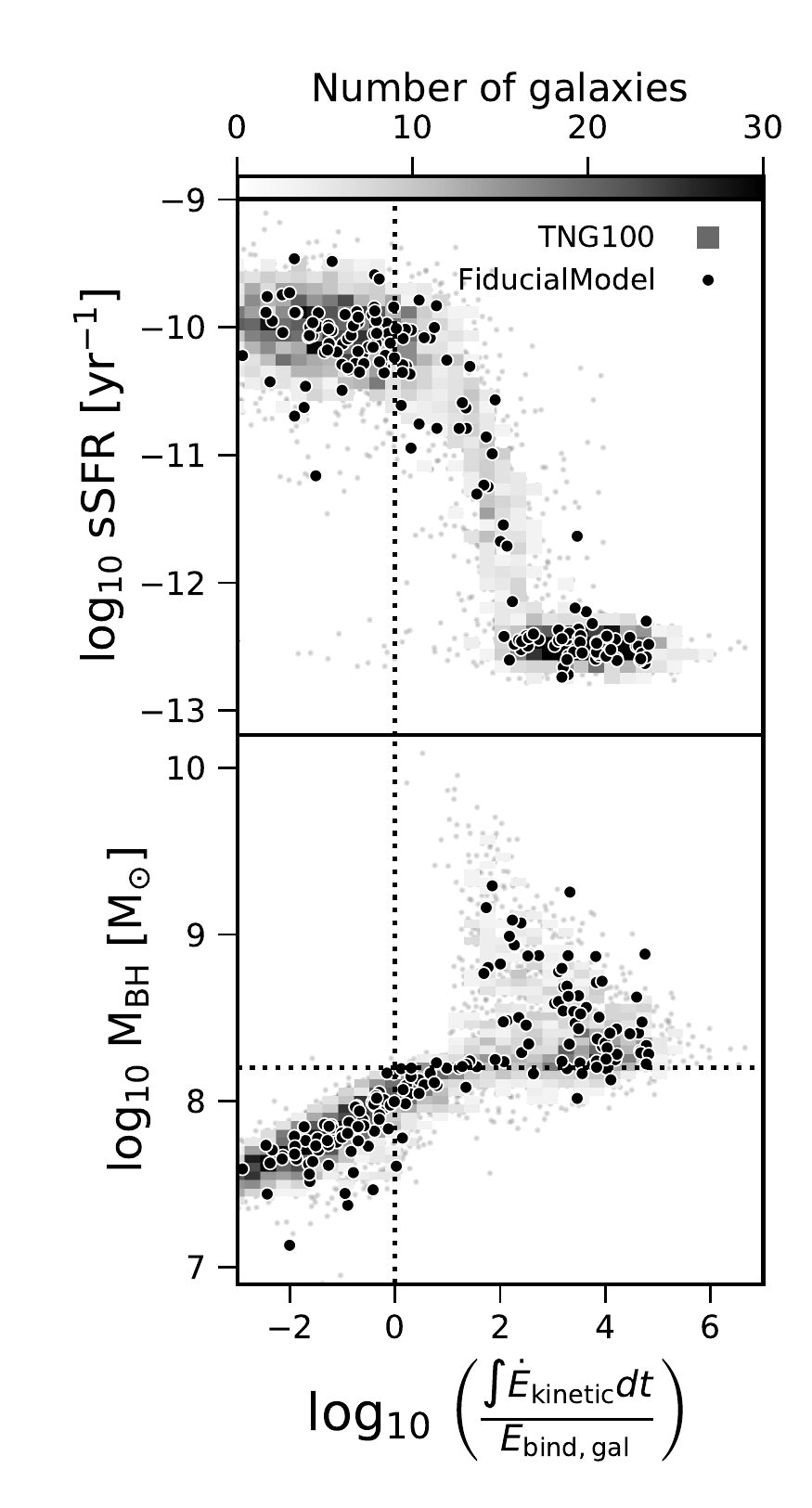}
\caption{sSFR (top) and M$_{\rm{BH}}$ (bottom) as a function of the ratio between the cumulative kinetic energy released from black hole feedback, $\int \dot{E}_{\rm{kinetic}} dt$, and the binding energy of the gas within twice the stellar half mass radius, $E_{\rm{bind, gal}}$. The vertical dotted line shows where these two energies equal. The grayscale heatmap shows the distribution of galaxies in TNG100 and the black points show galaxies in the FiducialModel simulation at $z = 0$. The sSFR decreases once the cumulative energy from black hole-driven winds exceeds the binding energy of gas in the galaxy. This drop in sSFR occurs when M$_{\rm{BH}}$ exceeds $10^{8.2}$ M$_{\odot}$ shown as a horizontal dotted line in the bottom panel.}
\label{fig:energies}
\end{figure}

\subsection{Overcoming gravitational binding energies}
\label{sec:phenom}

The results from Figure~\ref{fig:gasdistcomp} strongly suggest that black hole-driven kinetic winds drive gas out of the galaxy, producing a form of ejective feedback. Therefore, we choose to test whether the gravitational binding energy of the gas within the galaxy (defined in Section~\ref{sec:definitions}) can be a useful parameter for characterizing quiescence in the TNG model.

We compare this value to the time-integrated amount of black hole-driven wind energy that has been released into the gas particles near the black hole at each time-step, $\int \dot{E}_{\rm{kinetic}} dt$\footnote{This value is tracked for each black hole in the simulation and given as an output at each snapshot.}. \citet{dnps2018} and \citet{wsp2018} have demonstrated that quiescent galaxies have released more of this cumulative black hole wind energy relative to star-forming galaxies.

The top panel of Figure~\ref{fig:energies} shows the sSFR as a function of the ratio between the cumulative black hole wind energy, $\int \dot{E}_{\rm{kinetic}} dt$, and the gravitational binding energy of gas within the galaxy, $E_{\rm{bind, gal}}$, in the TNG100 (gray heatmap) and FiducialModel (black points) simulations at $z = 0$. We find that star-forming galaxies lie to the left of the vertical dotted line, showing that they exhibit cumulative black hole wind energies that fall below the binding energy of the gas. Galaxies with intermediate sSFRs between sSFR $\sim10^{-11} - 10^{-12}$ yr$^{-1}$ exhibit cumulative black hole wind energies that exceed the binding energy of the gas (to the right of the vertical dotted line) by up to a factor of 100. Above an energy ratio of 100, galaxies host very low sSFRs shown as upper limits at $\sim10^{-12.5}$ yr$^{-1}$.

It is important to note that the binding energy of the gas decreases as gas leaves the system since the total mass of the system decreases and there is less gas to push out. The high ratios of cumulative black hole wind energy to gravitational binding energy exceeding a factor of 1000 tend to have very low gas masses and therefore low binding energies. While this is true, we verify that sSFR correlates only slightly with the binding energy of gas within the galaxy, and that the cumulative black hole wind energy drives most of the correlation seen in the top panel of Figure~\ref{fig:energies}. The fact that sSFR begins to drop when these two energies equal indicates that gas is being gravitationally unbound from the central galaxy and pushed into the circumgalactic medium by black hole-driven kinetic winds.

The bottom panel of Figure~\ref{fig:energies} shows that galaxies with cumulative black hole wind energies lower than the binding energy of the gas have M$_{\rm{BH}} \lesssim10^{8.2}$ M$_{\odot}$, whereas those with higher ratios have M$_{\rm{BH}} \gtrsim10^{8.2}$ M$_{\odot}$. This indicates that black hole winds are effective at removing gas and producing quiescence for a majority of galaxies once the black hole exceeds the threshold mass at $\sim10^{8.2}$ M$_{\odot}$ (shown as a horizontal dotted line). 

We also note that the most massive black holes (M$_{\rm{BH}} \gtrsim 10^{9}$ M$_{\odot}$) tend to have lower ratios of black hole wind energy to binding energy, whereas less massive black holes (with M$_{\rm{BH}}$ still $\gtrsim10^{8.2}$ M$_{\odot}$) can have much higher ratios. This is likely due to the fact that more massive black holes live in more massive galaxies and haloes where the gravitational potential is much deeper. As such, many of the galaxies with intermediate sSFRs (between $\sim10^{-11} - 10^{-12}$ yr$^{-1}$) are hosted by some of the most massive haloes, whereas those with very low sSFRs (shown as upper limits at $\sim10^{-12.5}$) are those in lower mass haloes (with M$_{\rm{BH}}$ still $\gtrsim10^{8.2}$ M$_{\odot}$). We expand on this result in the following section. 

\subsection{The effects of black hole-driven kinetic feedback on the interstellar and circumgalactic media}
\label{sec:icgm}

\begin{figure}
\centering
\epsfxsize=7.9cm
\epsfbox{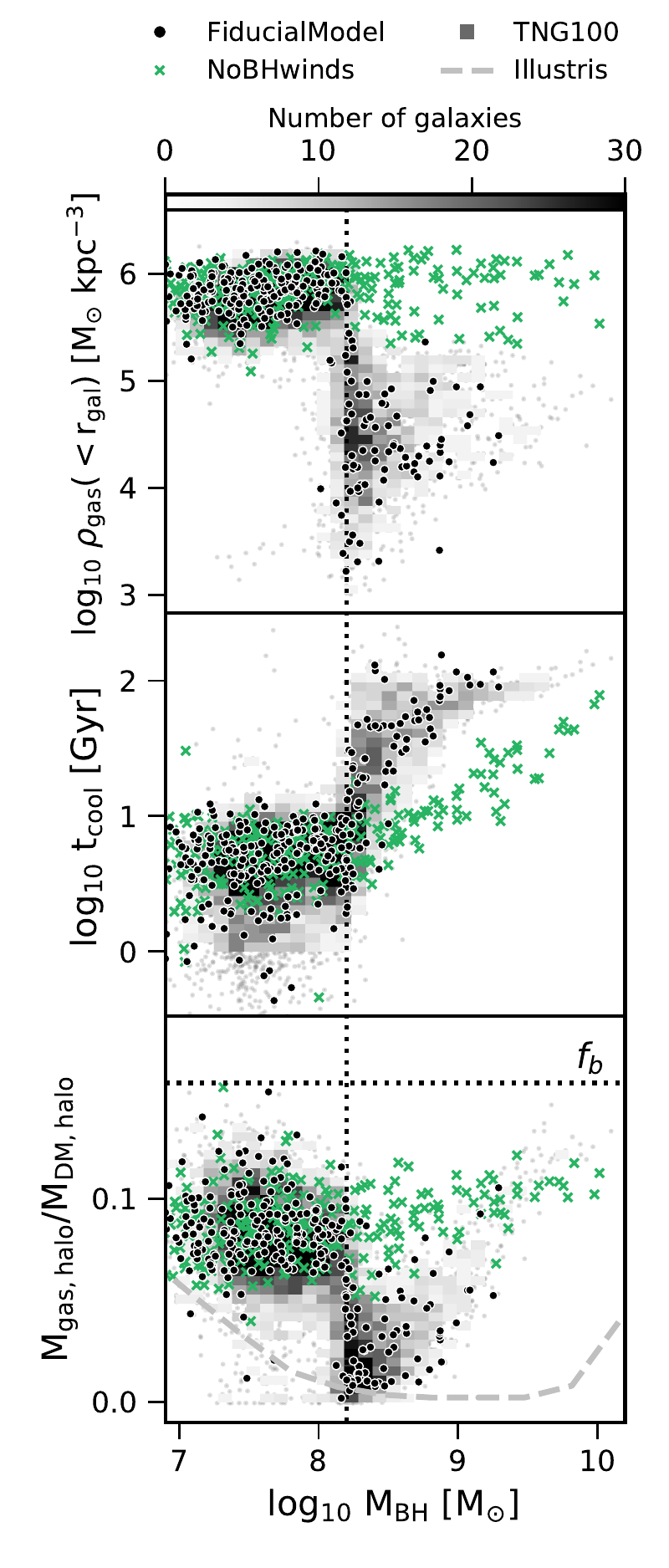}
\caption{The average gas density within the galaxy radius (top panel), the average cooling time of the halo gas (middle panel), and the ratio of halo gas mass to dark matter halo mass (bottom panel) as a function of M$_{\rm{BH}}$ for galaxies in the TNG100 (grayscale heatmap), FiducialModel (black), and NoBHwinds (green) simulations at $z = 0$. Including black hole winds in TNG abruptly lowers the density of gas within the galaxy, increases the cooling time of the halo gas, and reduces the amount of halo gas for galaxies with M$_{\rm{BH}} \gtrsim 10^{8.2}$ M$_{\odot}$. The most massive black holes live in the most massive haloes, where affecting gas kinetically is more difficult due to the halo's deep potential well. As a result, the galaxies whose SFRs are most affected by kinetic mode feedback are the least massive galaxies that have M$_{\rm{BH}} \gtrsim 10^{8.2}$ M$_{\odot}$.}
\label{fig:modelcomp}
\end{figure}

The bottom panel of Figure~\ref{fig:energies} indicates that there exists a M$_{\rm{BH}}$ threshold for quiescence (also see Figure 5 in \citealt{wsp2018}). In this section, we explicitly explore the gas properties of galaxies in TNG as a function of M$_{\rm{BH}}$. Figure~\ref{fig:modelcomp} shows the average gas density within the galaxy (top panel), the average cooling time of the entire gas halo\footnote{We calculate this value excluding star-forming cells. Refer to the footnote in Section~\ref{sec:sam}.} (middle panel), and the ratio of total halo gas mass to dark matter halo mass (bottom panel) as a function of M$_{\rm{BH}}$ for galaxies in the NoBHwinds (green crosses) and FiducialModel (black points) simulations at $z = 0$. The distribution of galaxies in TNG100 is shown as a grayscale heatmap.

Without kinetic winds (green crosses), the average gas density within twice the stellar half mass radius is similar across galaxies with different M$_{\rm{BH}}$ (top panel). We verify that this is because galaxy gas masses correlate closely with their radii when they are star-forming. The average cooling time of these galaxies' gas haloes is a fairly flat function of M$_{\rm{BH}}$ at low masses and gradually rises for more massive black holes that live in more massive haloes where cooling becomes less efficient (middle panel). Additionally, the total halo gas masses of these galaxies correlate with dark matter halo mass, producing a scattered, flat relation as a function of black hole mass (bottom panel).

Introducing kinetic winds significantly alters these gas properties for galaxies above the M$_{\rm{BH}}$ threshold for quiescence at $\sim10^{8.2}$ M$_{\odot}$, as is seen in all three panels of Figure~\ref{fig:modelcomp} (black points and grayscale heatmap). Once feedback from kinetic winds becomes effective, there is a sharp decrease in the average gas density within galaxies, a sharp increase in the average cooling time of halo gas, and a sharp decrease in the fractions of halo gas mass to dark matter halo mass compared to the NoBHwinds model.

%

%
\begin{figure}
\centering
\epsfxsize=8.3cm
\epsfbox{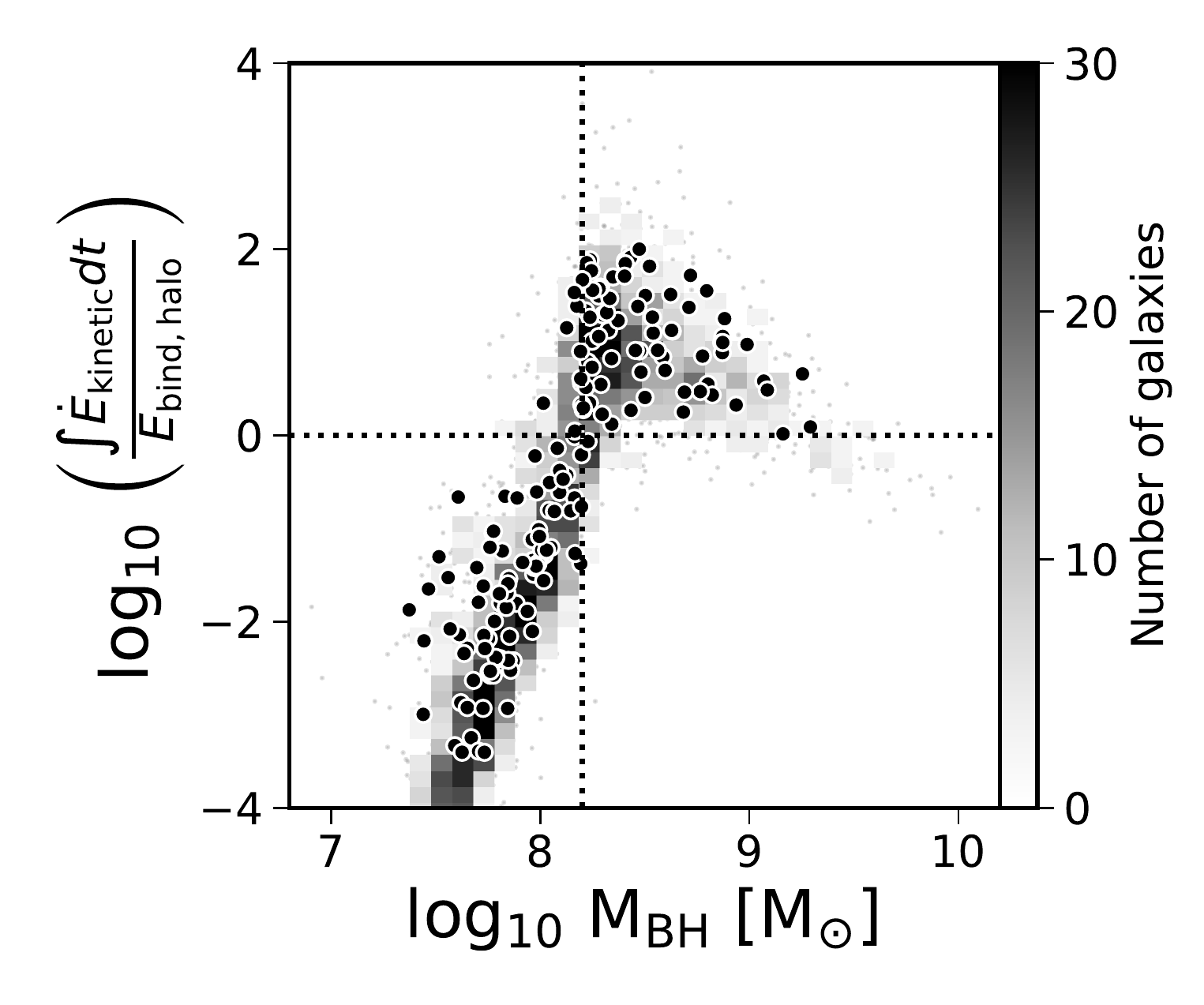}
\caption{The ratio between the cumulative energy from black hole winds, $\int \dot{E}_{\rm{kinetic}} dt$, and the binding energy of gas within the \textit{halo} radius, $E_{\rm{bind, halo}}$ as a function of M$_{\rm{BH}}$ for the TNG100 (grayscale heatmap) and FiducialModel (black points) simulations at $z = 0$. The horizontal dotted line indicates where the black hole wind energy equals the halo gas binding energy and the vertical dotted line indicates the M$_{\rm{BH}}$ threshold for quiescence (see Section~\ref{sec:phenom}). The black hole wind energy exceeds the halo gas binding energy by the largest amount just above the M$_{\rm{BH}}$ threshold for quiescence.}
\label{fig:halobind}
\end{figure}

\begin{figure*}
\centering
\epsfxsize=17.5cm
\epsfbox{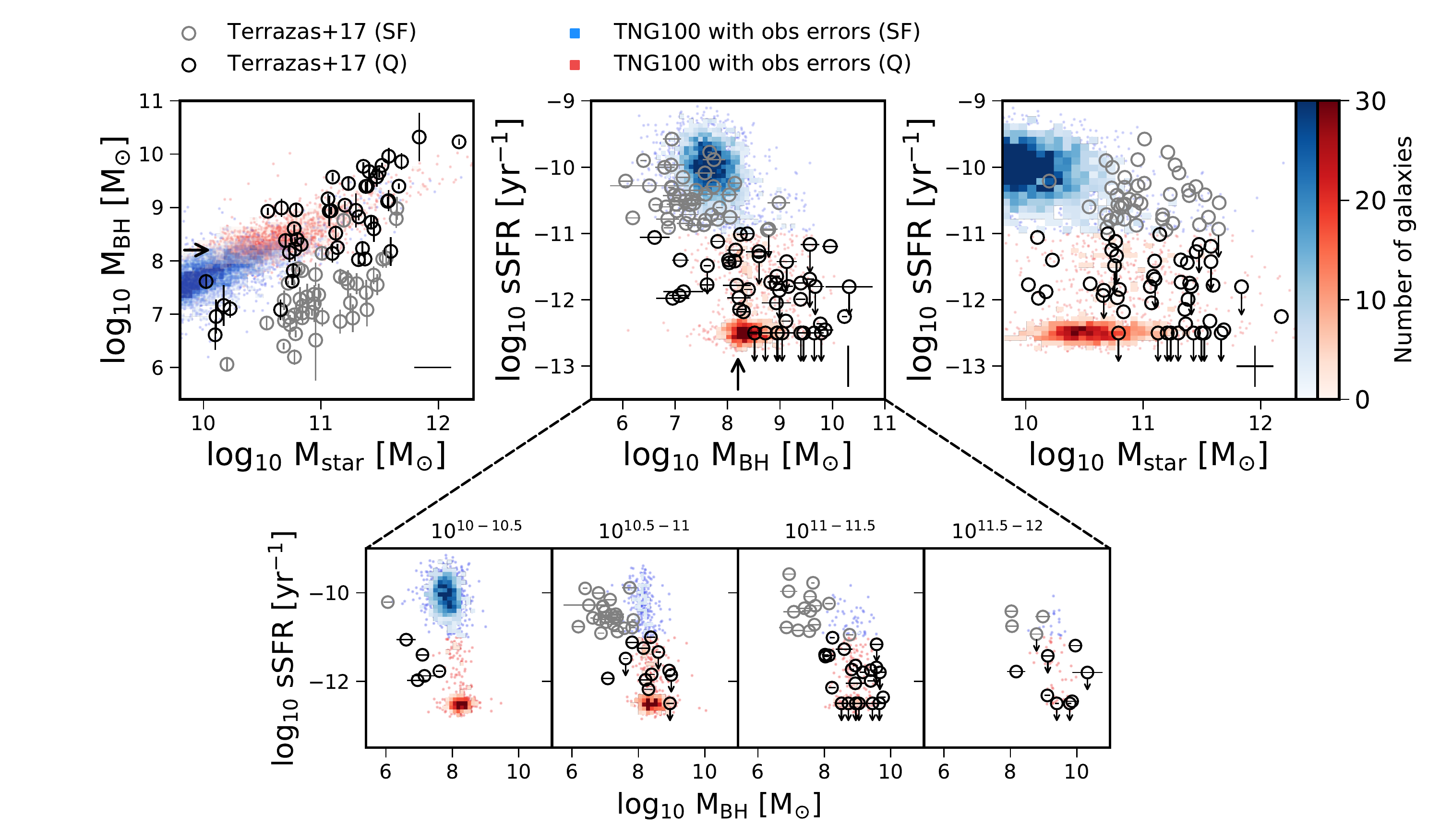}
\caption{\textit{Top panels:} The M$_{\rm{BH}}$-M$_{\rm{star}}$ (left), sSFR-M$_{\rm{BH}}$ (centre), and sSFR-M$_{\rm{star}}$ (right) parameter spaces for TNG100 galaxies in blue and red heatmaps and the observational sample in gray and black circles, for star-forming and quiescent galaxies, respectively, at $z = 0$. An uncertainty of 0.15 dex is added to sSFR and M$_{\rm{star}}$ values for TNG100 galaxies. Observational uncertainties for M$_{\rm{BH}}$ in TNG100 are taken into account by convolving them with the errors from the T17 data. \textit{Bottom panels:} sSFR as a function of M$_{\rm{BH}}$ in bins of M$_{\rm{star}}$ for TNG100 galaxies and the T17 observational sample.}
\label{fig:3Dcube}
\end{figure*}

In the bottom panel we also show the median values from the original Illustris model with a dashed gray line. This demonstrates that while TNG retains more gas in massive haloes on average, it still removes large quantities of gas from within the halo's radius once black hole winds are produced at the M$_{\rm{BH}}$ threshold for quiescence. The lowest halo gas fractions (with M$_{\rm{gas, halo}}$/M$_{\rm{DM, halo}} \sim 0$) are seen in galaxies with M$_{\rm{BH}}$ near the threshold mass at $\sim 10^{8.2}$ M$_{\odot}$. The gas fraction increases with M$_{\rm{BH}}$ thereafter. This is likely because M$_{\rm{BH}}$ correlates closely with M$_{\rm{star}}$ (as we will show in Figure~\ref{fig:3Dcube}), and therefore M$_{\rm{DM,halo}}$, for TNG galaxies. The feedback from black holes in more massive galaxies must overcome a larger potential well in order to expel gas from the galaxy's halo.

To illustrate this, Figure~\ref{fig:halobind} shows the ratio between the cumulative energy from black hole winds and the gravitational binding energy of the \textit{halo gas} as a function of M$_{\rm{BH}}$. The binding energy of the halo gas is calculated by replacing $r_{\rm{gal}}$ with $r_{\rm{halo}}$ in Eqn.~\ref{eqn:be} (see Section~\ref{sec:definitions}). The horizontal dotted line shows where the black hole kinetic wind energy equals the halo gas binding energy. Galaxies hosting black holes just over the M$_{\rm{BH}}$ threshold for quiescence (vertical dotted line) have the highest ratios of black hole wind energy to halo gas binding energy. This coincides with the lowest gas mass fractions in the lower panel of Figure~\ref{fig:modelcomp}. This indicates that a significant fraction of gas is evacuated from many of the haloes once black hole wind energies exceed not only the binding energy of the gas in the galaxy (Figure~\ref{fig:energies}) but also of the gas in the halo.

Progressively higher mass black holes live in more massive haloes with deeper gravitational potentials. In Figure~\ref{fig:halobind} as well as in the lower panel of Figure~\ref{fig:energies}, the ratio between the energy from black hole winds and the binding energy of gas in both the galaxy and the halo decreases for more massive black holes. The decrease in this ratio with black hole mass coincides with an increase in the halo gas fraction in the lower panel of Figure~\ref{fig:modelcomp}. This behavior indicates that black hole winds are less effective at removing gas at higher mass regimes.

\section{Comparisons to Observations}
\label{sec:compobs}

In this section we evaluate the extent to which the kinetic wind model in TNG can produce results which agree with currently available observational correlations related to quiescence. Here we use the observational diagnostics and data from \citet{tbh2016b, tbw2017} in order to examine and compare the relationship between M$_{\rm{star}}$, M$_{\rm{BH}}$, and sSFR in TNG. We aim to link the phenomenological description of quiescence we put forth for this model in Section~\ref{sec:quiescence} and~\ref{sec:physics} to these observable properties of simulated galaxies since these properties are causally tied to one another in this model as a result of the black hole feedback prescriptions.

%

\subsection{Observational sample}
\label{sec:obs}

The 91 galaxies in \citet[][hereafter T17]{tbh2016b, tbw2017} represent a diverse collection of local galaxies (distances within $\sim$150 Mpc) with various morphologies, SFRs, colours, and environments (isolated, group, or cluster) with M$_{\rm{star}} > 10^{10}$ M$_{\odot}$. The sample represents central galaxies -- defined as the most massive galaxy within $\sim$1 Mpc -- that have a dynamical M$_{\rm{BH}}$ measurement using stellar dynamics (45 galaxies), gas dynamics (15), masers (12), or reverberation mapping (18). Black hole mass measurements were taken from \citet{soe2016} and \citet{v2016}.

T17 measures the stellar masses using 2MASS $K_{\rm{s}}$-band luminosities \citep{hmm2012} with a single M$_{\rm{star}}$/$L_{\rm{K_{\rm{s}}}}$ ratio of 0.75 since mass-to-light ratios for $L_{\rm{K_{\rm{s}}}}$ do not vary substantially \citep{bd2001}. Their scatter in mass-to-light ratios are $\sim$0.15 dex, thus they assume the same value for their uncertainty. SFRs are measured following \citet{ke2012} where T17 use IRAS far-infrared flux measurements to estimate a total infrared luminosity. This SFR estimate is sensitive to star formation during the past $\sim$100 Myr. SFRs are assumed to have an uncertainty of 0.15 dex following \citet{b2003}. Any SFRs with no far-infrared detection, fluxes below the detection limits of IRAS, or detections below sSFR $< 10^{-12.5}$ yr$^{-1}$ are taken as upper limits. We define star-forming and quiescent galaxies to be those with sSFRs above or below $10^{-11}$ yr$^{-1}$, respectively, just as we do with TNG galaxies. 

We note that these data are not representative of the entire galaxy population at $z = 0$ due to the requirement that their M$_{\rm{BH}}$ be large enough and their host galaxies close enough to detect their gravitational sphere of influence. The various detection methods used for this sample make it difficult to fully understand the biases associated with the sample. For example, black hole masses detected with reverberation mapping or masers are more likely to be in star-forming galaxies where there is enough gas to produce the emission required to use these measurement techniques. Black holes measured with stellar dynamics, however, are likely in systems that are close enough to resolve the gravitational sphere of influence and that have low enough gas masses in their nuclear regions to allow starlight to dominate. These factors make it difficult to compare these observations directly with the simulation results, since the selection function and bias for the observational sample with regards to its sSFR, M$_{\rm{star}}$, and M$_{\rm{BH}}$ distributions are not well understood.

One may decide to use a sample with proxies instead, with the advantage that a complete, representative sample may be obtained. However, proxies for M$_{\rm{BH}}$ such as velocity dispersion, bulge mass, or S\'ersic index exhibit intrinsic scatter that likely indicates differences in the physics that set these properties \citep[e.g.,][]{grg2009, bcc2012, sbs2016, tbh2016b, mbv2016, mbr2018}. Using proxies for M$_{\rm{BH}}$ would also introduce unknown uncertainties that likely depend on M$_{\rm{star}}$. While the number density of galaxies at a given sSFR, M$_{\rm{star}}$, and M$_{\rm{BH}}$ are uncertain for this sample, our primary concern for this study is accuracy. Using dynamical black hole masses also ensures that the measurements for M$_{\rm{BH}}$ and M$_{\rm{star}}$ are independent from one another. Finally, while the sample may not be representative quantitatively, it contains a large diversity of galaxy properties ranging 4 orders of magnitude in sSFR, 2 orders of magnitude in M$_{\rm{star}}$, and 4 orders of magnitude in M$_{\rm{BH}}$. This results from the variety of M$_{\rm{BH}}$ measurement techniques that sample distinct parts of the whole galaxy population. As such, we opt to use high-quality M$_{\rm{BH}}$ estimates instead of proxies in order to reliably understand how galaxy properties correlate with M$_{\rm{BH}}$.

\subsection{Results}
\label{sec:results}

Figure~\ref{fig:3Dcube} shows the M$_{\rm{BH}}$-M$_{\rm{star}}$ (top left), sSFR-M$_{\rm{BH}}$ (top centre), and sSFR-M$_{\rm{star}}$ (top right) parameter spaces for the TNG100 simulation at $z = 0$ (star-forming: blue heatmap, quiescent: red heatmap), and the observational data from T17 (star-forming: gray circles, quiescent: black circles). On the lower right corner of each panel we indicate the 0.15 dex uncertainties on the sSFR and M$_{\rm{star}}$ of the observational sample. We convolve the simulation quantities for sSFR and M$_{\rm{star}}$ with these uncertainties in order to more fairly compare the observations with the simulation results. Similarly, we convolve the simulation quantities for M$_{\rm{BH}}$ with uncertainties from the observed M$_{\rm{BH}}$ sample. The bottom panels show sSFR as a function of M$_{\rm{BH}}$ in bins of M$_{\rm{star}}$, whose value is denoted above each plot.

The top leftmost panel of Figure~\ref{fig:3Dcube} shows that TNG produces a tight M$_{\rm{BH}}$-M$_{\rm{star}}$ relation, even with added observational uncertainties that increase the scatter. While TNG is in good agreement with the qualitative result that quiescent galaxies host more massive black holes, quantitatively the distributions of star-forming and quiescent galaxies differ between TNG and the observations. Namely, the observational data show significant scatter in M$_{\rm{BH}}$ as a function of \textit{total} M$_{\rm{star}}$. We note that the relation shown in the top leftmost panel of Figure~\ref{fig:3Dcube} differs from canonical M$_{\rm{BH}}$-galaxy scaling relation studies in that a \textit{total} M$_{\rm{star}}$ is preferred in this work over a \textit{bulge-only} M$_{\rm{star}}$.

The scatter in the observed M$_{\rm{BH}}$-M$_{\rm{star}}$ relation correlates well with the sSFR of these galaxies, where quiescent galaxies host more massive black holes than star-forming galaxies. TNG does not easily produce galaxies with the M$_{\rm{BH}}$ and M$_{\rm{star}}$ demographics of star-forming galaxies in the T17 sample. For example, the Milky Way (included in the T17 sample) with M$_{\rm{star}} \approx 10^{10.7}$ M$_{\odot}$ and M$_{\rm{BH}} \approx 10^{6.7}$ M$_{\odot}$ is not represented in TNG100. Instead, TNG predicts that the distribution of star-forming galaxies lies far to the left of where they lie in the observational sample.

Additionally, the boundary between star-forming and quiescent galaxies on the top leftmost plot showing the M$_{\rm{BH}}$-M$_{\rm{star}}$ relation for TNG is flat, representing a M$_{\rm{BH}}$ threshold for quiescence at $\sim10^{8.2}$ M$_{\odot}$, as was discussed in Section~\ref{sec:physics}. The slope of the boundary in the observational data is positive with a value of $\sim1$. T17 note that this slope represents lines of constant M$_{\rm{BH}}$/M$_{\rm{star}}$ and that the sSFR decreases perpendicular to these lines moving towards more massive black holes. As such, while quiescence does correlate with M$_{\rm{BH}}$ in the observational sample, a threshold for quiescence does not exist at a particular M$_{\rm{BH}}$ value as it does in TNG.

We show this more explicitly in the top middle panel of Figure~\ref{fig:3Dcube} where sSFR is shown as a function of M$_{\rm{BH}}$. There is a sharp transition between star-forming and quiescent galaxies at the M$_{\rm{BH}}$ threshold for quiescence at $\sim10^{8.2}$ M$_{\odot}$ for TNG galaxies. Galaxies having very low sSFRs shown as upper limits at sSFR $\simeq 10^{-12.5}$ yr$^{-1}$, make up 73\% of the total quiescent galaxy population in TNG. This produces a strong apparent bimodality in sSFR for the simulated galaxies (in agreement with TNG results shown and discussed in \citealt{dpn2019}) as a result of the black hole feedback model at low accretion rates.

The observational data show a scattered, more gradual decrease in sSFR as a function of M$_{\rm{BH}}$ compared to TNG. Namely, there is no indication in the observational data that there is a sharp M$_{\rm{BH}}$ threshold where galaxies mostly exhibit very low sSFRs ($\simeq 10^{-12.5}$ yr$^{-1}$). T17 measure significant, intermediate sSFR values ($\sim10^{-11} - 10^{-12}$ yr$^{-1}$) for many galaxies with massive black holes $\gtrsim10^{8.2}$ M$_{\odot}$. Empirically, the sSFR of the observational sample is a tighter and more smoothly declining function of M$_{\rm{BH}}$/M$_{\rm{star}}$ rather than M$_{\rm{BH}}$ alone, resulting in the scattered relationship between sSFR and M$_{\rm{BH}}$.

We note that low SFRs in early-type galaxies are notoriously difficult to measure \citep[e.g.][]{cby2011}. The sSFRs from T17 are measured using far-infrared IRAS luminosities in order to avoid contamination from possible AGN. They also used ultraviolet luminosities from GALEX to check the robustness of their values, finding consistent results using different techniques. Additionally, recent work indicates the existence of a population of galaxies with low but significant levels of star formation \citep{oag2016}, especially when measured in the infrared \citep{edw2017}. The morphologies and colours of galaxies at intermediate sSFRs in the T17 sample are also visually more disc-like and bluer compared to the more spheroidal and redder galaxies at lower sSFRs, providing further evidence for the robustness of the measured difference between galaxies at intermediate ($\simeq 10^{-11}-10^{-12}$ yr$^{-1}$) compared to low ($\simeq 10^{-12.5}$ yr$^{-1}$) sSFRs in the T17 sample.

In order to more clearly show the relationship between sSFR and M$_{\rm{BH}}$/M$_{\rm{star}}$ in the observational sample, we plot sSFR as a function of M$_{\rm{BH}}$ in four bins of M$_{\rm{star}}$ in the bottom panels of Figure~\ref{fig:3Dcube}. These plots show a gradual decrease in sSFR as a function of M$_{\rm{BH}}$ in each M$_{\rm{star}}$ bin for the observational sample (see \citealt{tbw2017} for a full quantitative discussion on this empirical relationship). TNG adequately reproduces the M$_{\rm{BH}}$ distribution of the quiescent population in all but the lowest M$_{\rm{star}}$ bin. However, TNG and the observational data differ significantly for the star-forming population, where TNG predicts much higher M$_{\rm{BH}}$ values compared to the observational data at each M$_{\rm{star}}$ bin. Additionally, TNG shows a much narrower distribution of M$_{\rm{BH}}$ values at each M$_{\rm{star}}$ bin than in the observations due to the tightness of their M$_{\rm{BH}}$-M$_{\rm{star}}$ relation.


The top right plot shows the sSFR as a function of M$_{\rm{star}}$. The strong bimodality in TNG galaxies as a result of black hole winds is evident, particularly between M$_{\rm{star}} = 10^{10-11}$ M$_{\odot}$. Additionally, the most massive galaxies in TNG exhibit low yet significant amounts of sSFRs moreso than galaxies at lower masses. We discussed this in Section~\ref{sec:icgm}, where black hole wind energy was less effective against a more massive potential well where the gas can be more readily retained due to the higher binding energies in these galaxies. 

The sSFR-M$_{\rm{star}}$ parameter space also highlights the biases in the T17 sample. There is no star forming main sequence, as is seen in more representative samples of galaxies in this parameter space \citep[e.g.,][]{cvd2015}. However, this panel also shows how the sample spans a large range of M$_{\rm{star}}$ and sSFR.

%
\begin{figure}
\centering
\epsfxsize=7.8cm
\epsfbox{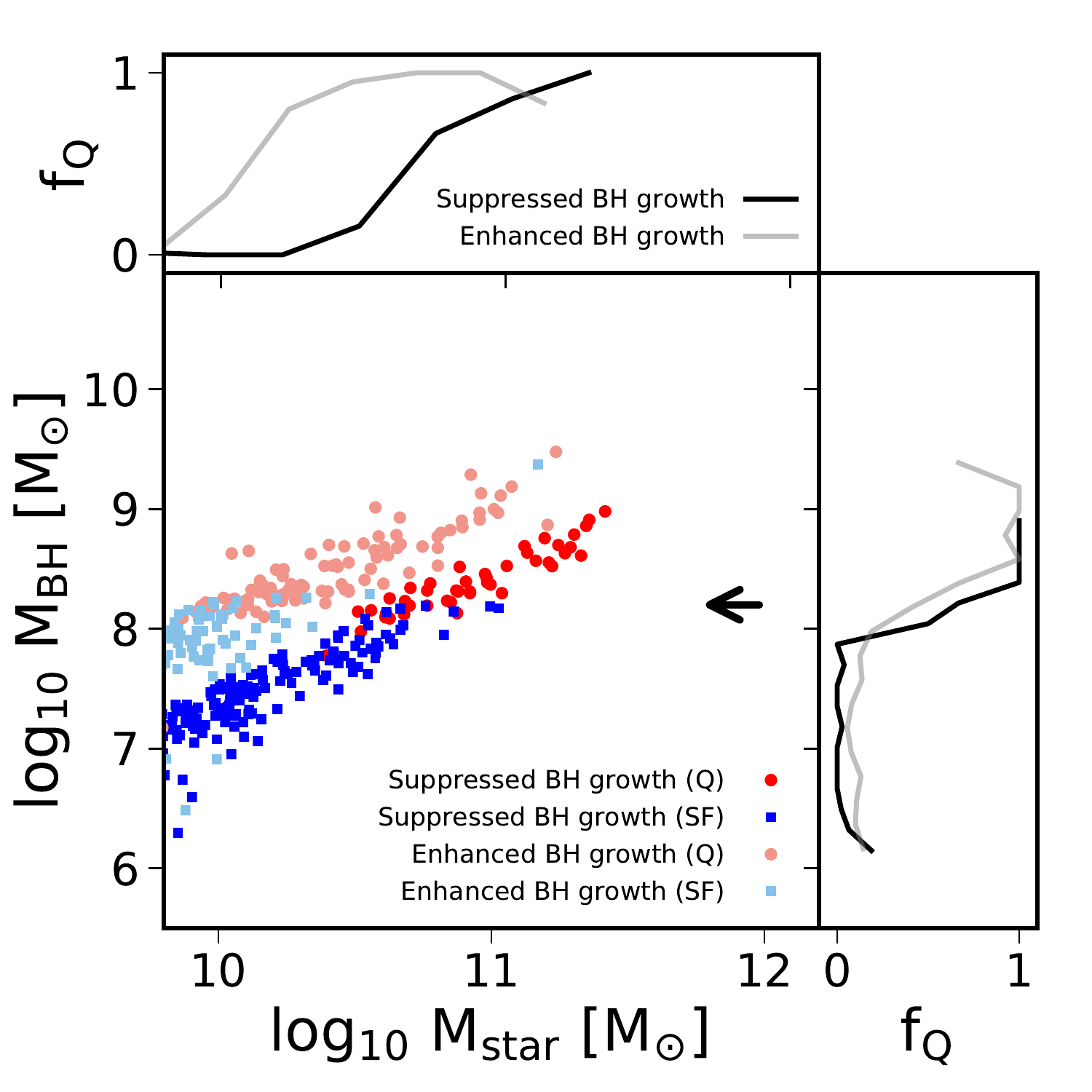}
\caption{The M$_{\rm{BH}}$-M$_{\rm{star}}$ relation at $z = 0$ for the LowThermEff (light blue and red points) and HighThermEff (dark red and blue points) model variations. The top and right panels show the quiescent fraction as a function of M$_{\rm{star}}$ and M$_{\rm{BH}}$, respectively, for the LowThermEff (gray) and HighThermEff (black) simulations. The main difference between the two models is the normalization of the M$_{\rm{BH}}$-M$_{\rm{star}}$ relation since thermal mode feedback primarily regulates M$_{\rm{BH}}$ growth. Both simulations have the same M$_{\rm{BH}}$ threshold for quiescence (black arrow) yet the change in normalization alters the quiescent fraction as a function of M$_{\rm{star}}$.}
\label{fig:norm}
\end{figure}

\begin{figure}
\centering
\epsfxsize=8.7cm
\epsfbox{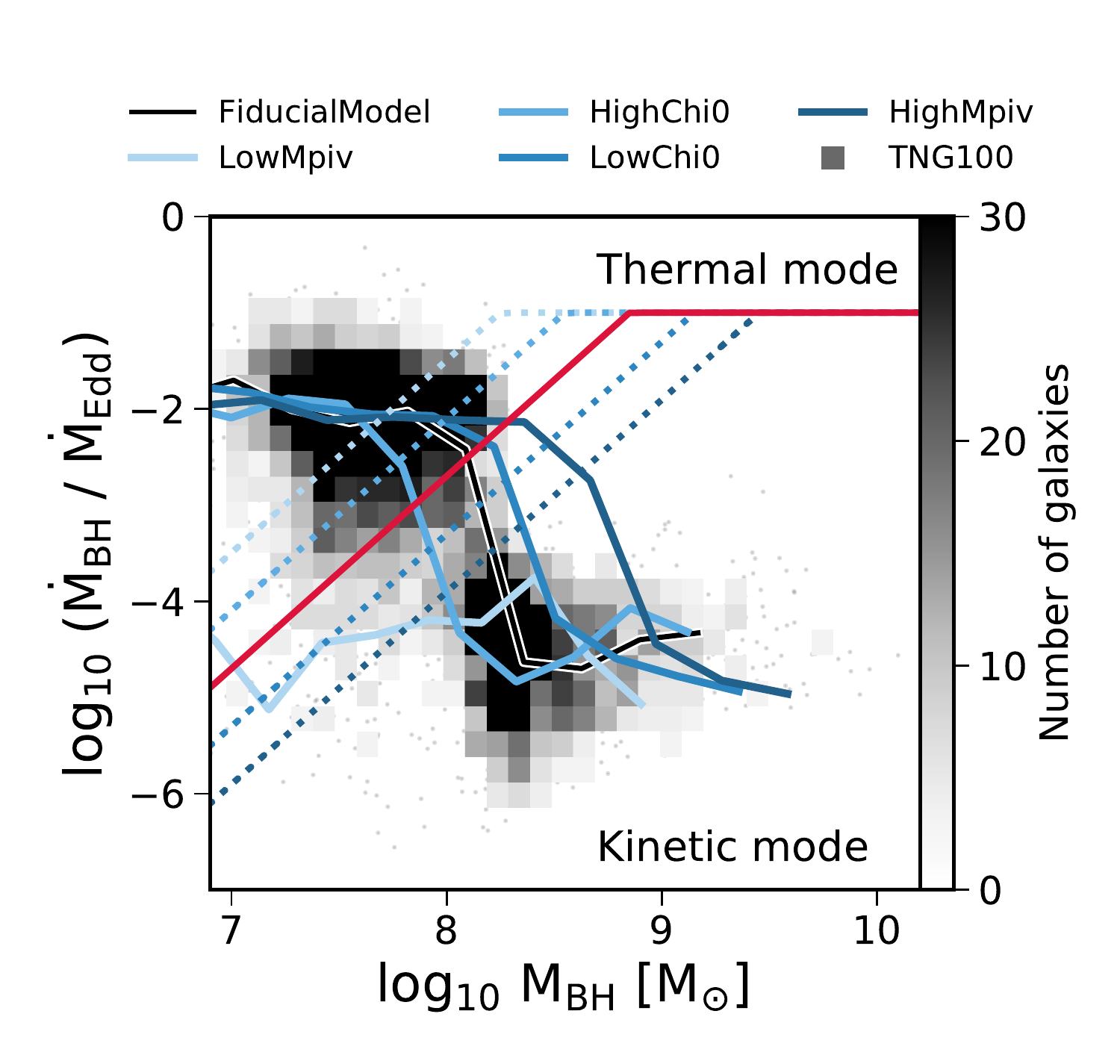}
\caption{The Eddington ratio as a function of M$_{\rm{BH}}$ for the TNG100 population of galaxies (grayscale heatmap) along with the Eddington ratio threshold that determines whether a black hole is producing kinetic or thermal feedback energy (red solid line) at $z = 0$. The median distributions of the LowMpiv, HighChi0, LowChi0, and HighMpiv model variations at $z = 0$ are also plotted (blue lines). These models alter the threshold described in Eqn.~\ref{eqn:chi} (shown as blue dotted lines).}
\label{fig:boundary}
\end{figure}

%
\begin{figure*}
\centering
\epsfxsize=4.3cm
\epsfbox{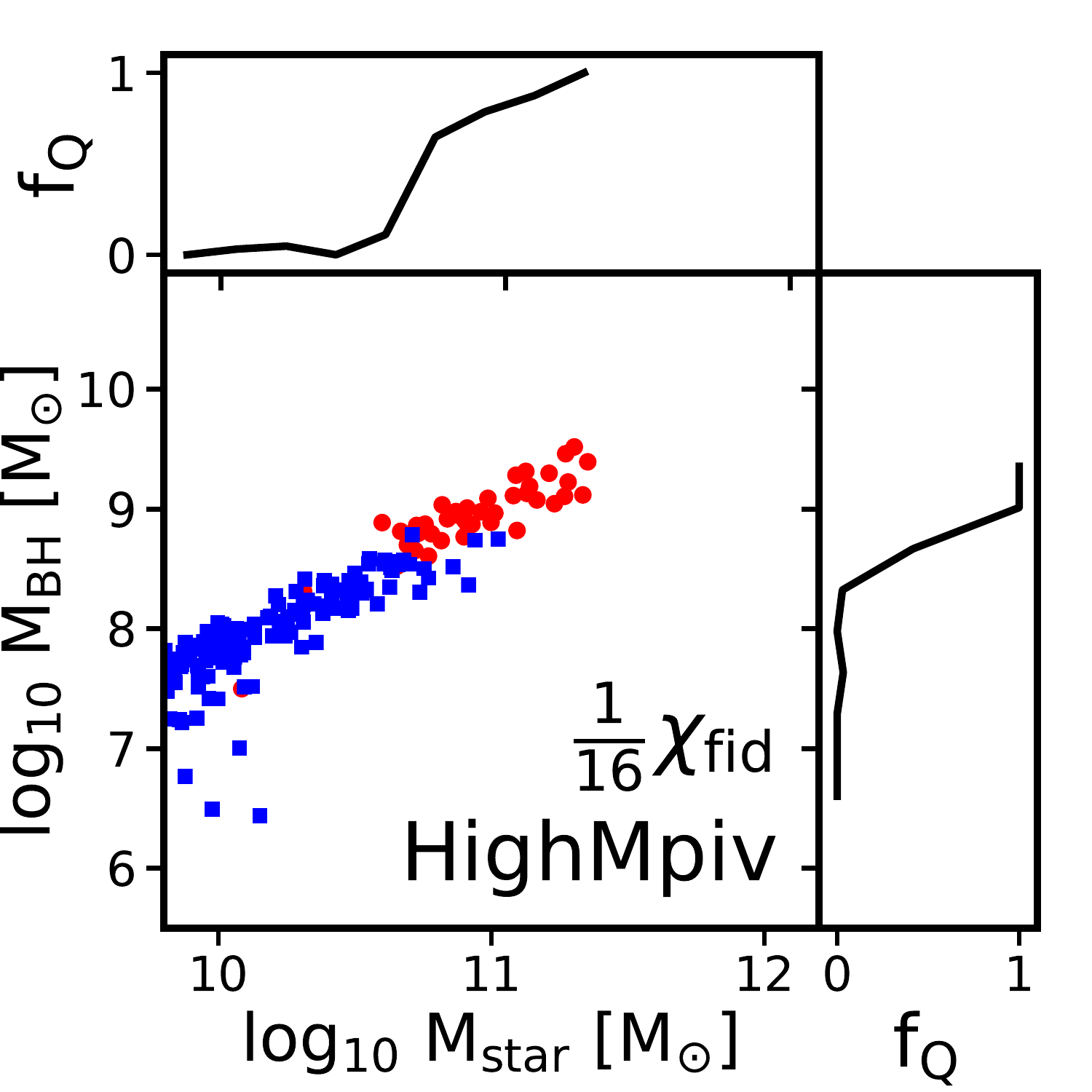}
\epsfxsize=4.3cm
\epsfbox{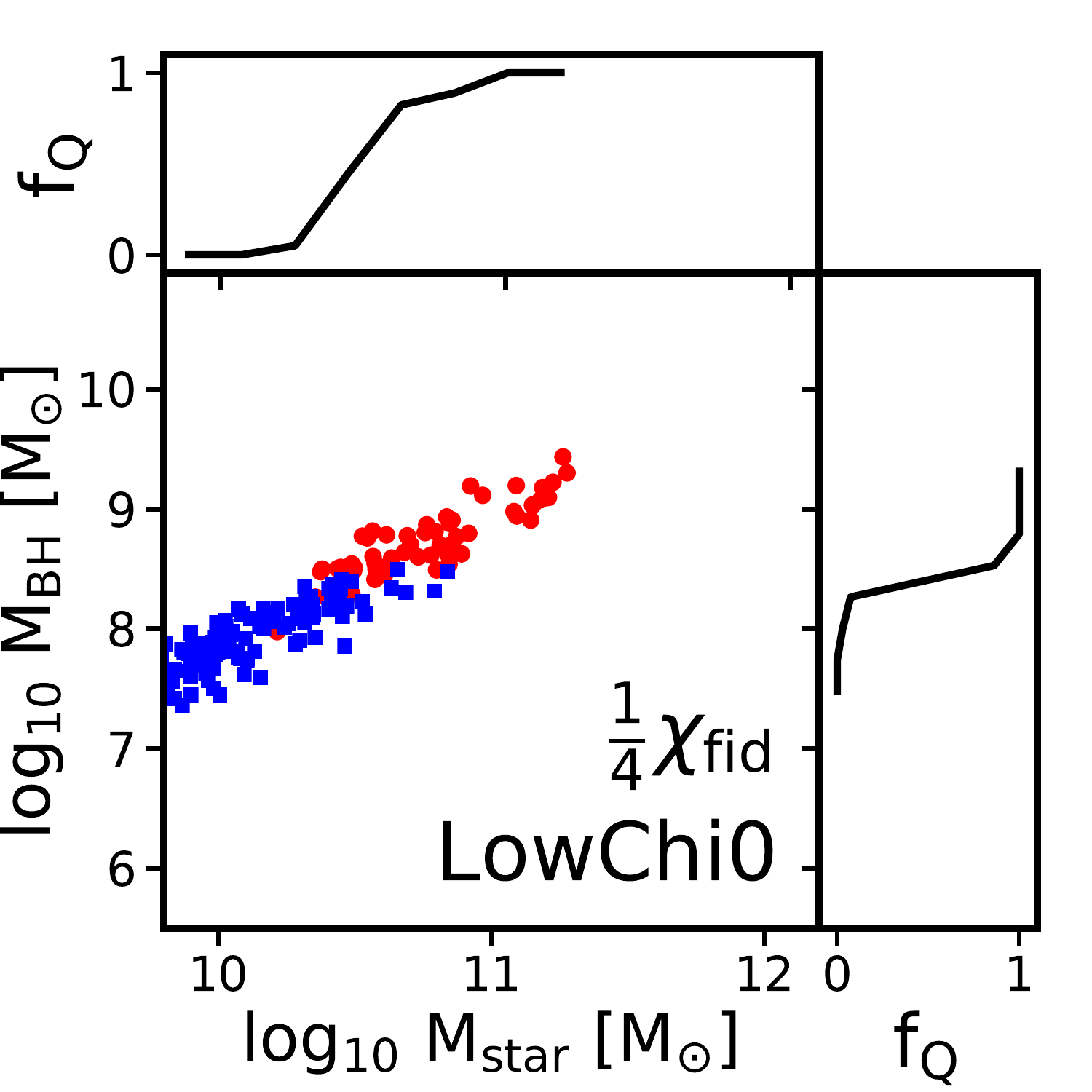}
\epsfxsize=4.3cm
\epsfbox{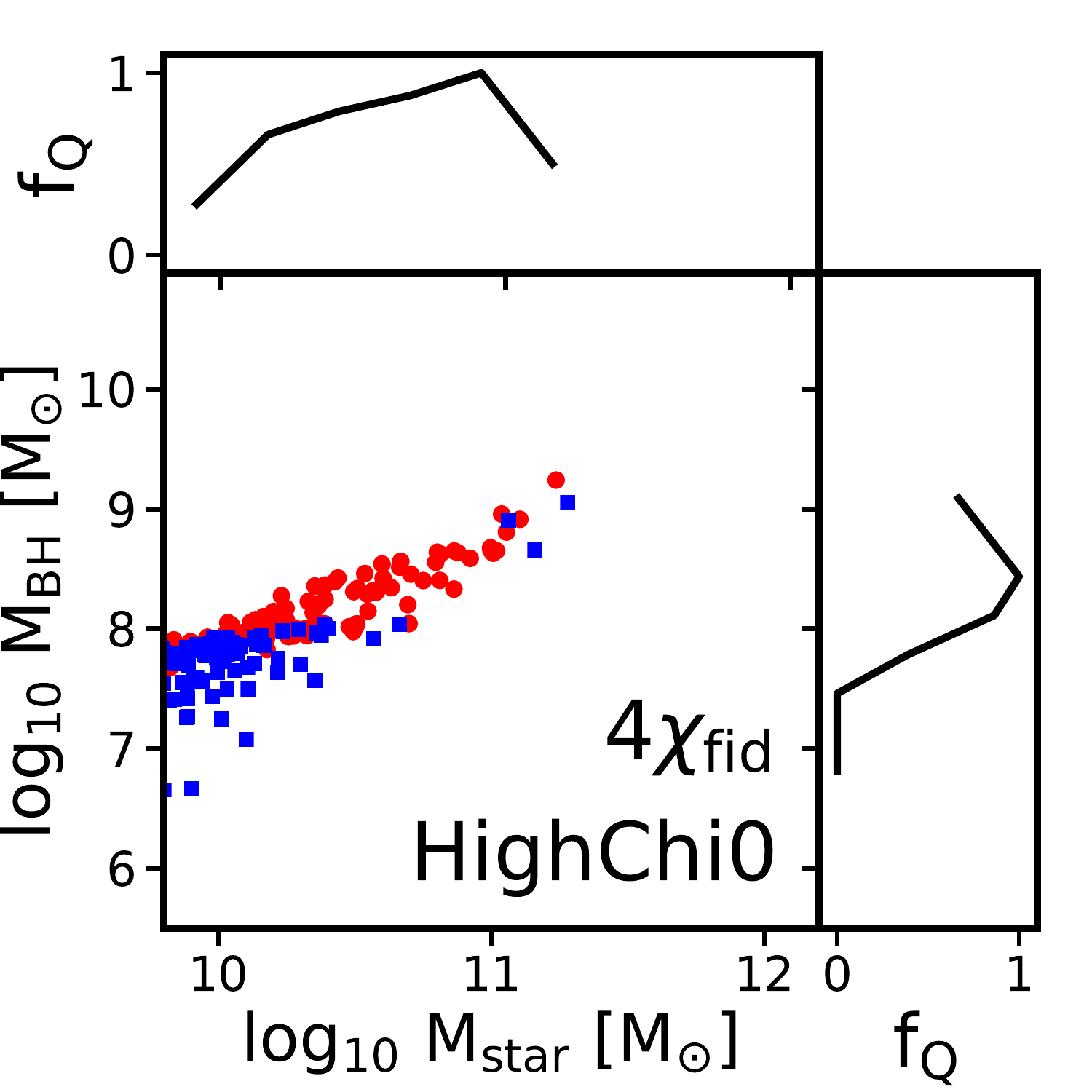}
\epsfxsize=4.3cm
\epsfbox{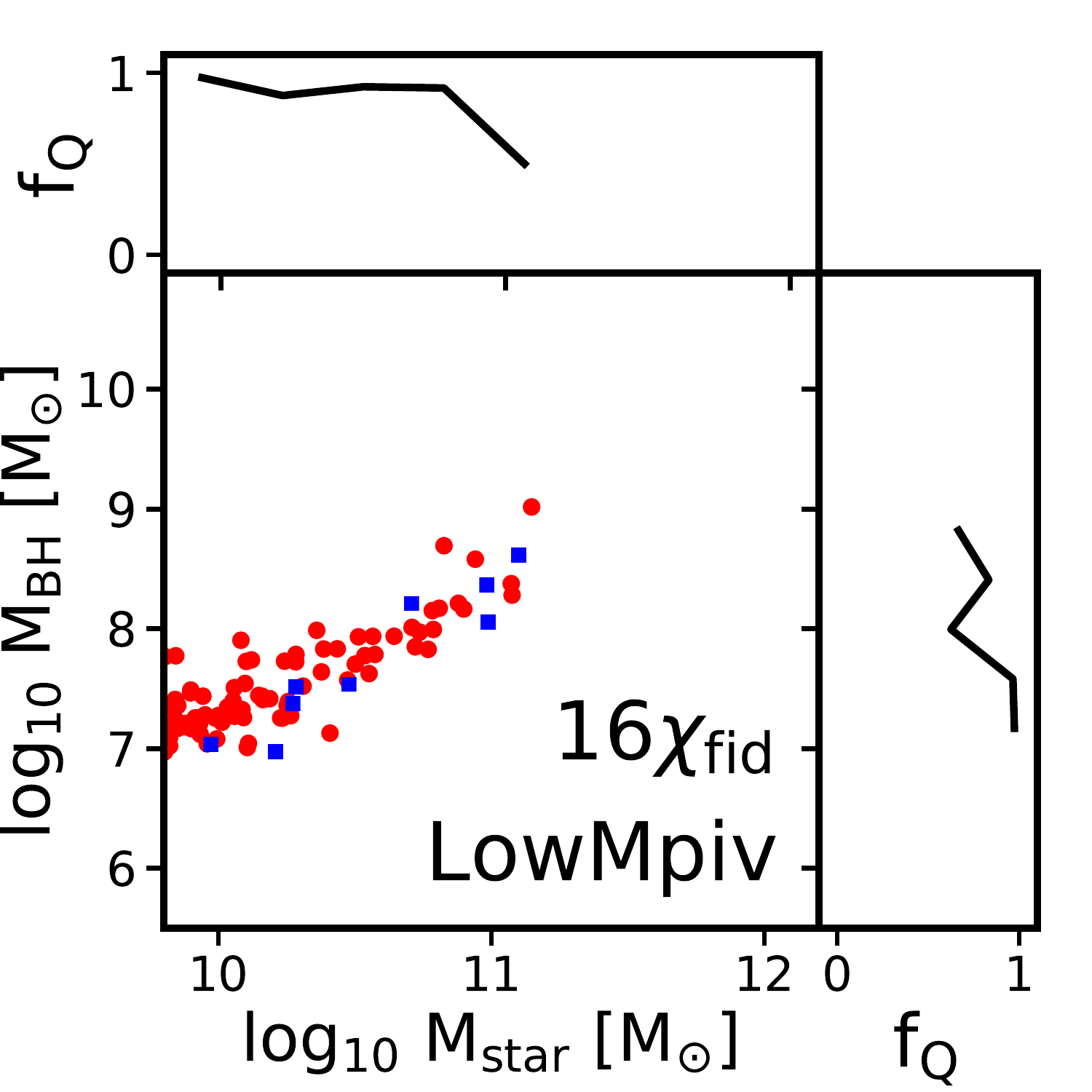}
\caption{The M$_{\rm{BH}}$-M$_{\rm{star}}$ relation for HighMpiv, LowChi0, HighChi0, and LowMpiv model variations at $z = 0$. The quiescent fraction as a function of M$_{\rm{star}}$ and M$_{\rm{BH}}$ is shown on top and to the right of each plot, respectively. As the Eddington ratio threshold between thermal and kinetic mode black hole feedback ($\chi$) increases in units of the fiducial model's value from left to right, the black hole mass at which galaxies are able to become quiescent decreases.}
\label{fig:mbhthreshold}
\end{figure*}

\section{Implications of the black hole feedback model on observational diagnostics}
\label{sec:implications}

In this section, we explore how the distribution of galaxies on the sSFR-M$_{\rm{star}}$-M$_{\rm{BH}}$ parameter space is sensitive to changes in the physics of quiescence using the TNG model variations.

\subsection{The coevolution of M$_{\rm{BH}}$ and M$_{\rm{star}}$}
\label{sec:coevol}

The strong correlation between M$_{\rm{BH}}$ and M$_{\rm{star}}$ paired with a M$_{\rm{BH}}$ threshold for quiescence produces a fairly narrow range of M$_{\rm{star}}$ where both quiescent and star-forming galaxies can coexist with comparable numbers in TNG. The quiescent fraction rises from 0.3 to 0.7 between M$_{\rm{star}} = 10^{10.25-10.55}$ M$_{\odot}$ (in general agreement with TNG results discussed in \citealt{dpn2019}), giving a 0.3 dex range in M$_{\rm{star}}$. As a consequence of these features, the distribution of star formation as a function of a galaxy's M$_{\rm{star}}$ is particularly sensitive to the normalization of the M$_{\rm{BH}}$-M$_{\rm{star}}$ relation.

We show this in Figure~\ref{fig:norm} where we plot the M$_{\rm{BH}}$-M$_{\rm{star}}$ relations at $z = 0$ for the LowThermEff (light blue and red) and HighThermEff (dark blue and red) model variations, which show two simulations where thermal mode feedback is less and more efficient, respectively. The blue and red colours indicate whether galaxies are star-forming or quiescent, i.e. above or below sSFR = $10^{-11}$ yr$^{-1}$. The top and right panels show the quiescent fraction as a function of M$_{\rm{star}}$ and M$_{\rm{BH}}$ for the LowThermEff (gray) and HighThermEff (black) simulations. 

In TNG, the thermal mode primarily regulates the growth of the black hole (see Fig. 7 in \citealt{wsp2018} and the discussion therein for a detailed description of M$_{\rm{BH}}$ growth in TNG during different accretion phases). A high efficiency of thermal mode feedback more readily suppresses a black hole's growth since it transfers thermal energy to the gas immediately surrounding it, reducing the amount of cold gas that can accrete. As a result, while the M$_{\rm{BH}}$ threshold for quiescence does not change as a result of different thermal mode efficiencies (see right hand panel of Figure~\ref{fig:norm}), the distribution of star formation amongst galaxies at different M$_{\rm{star}}$ (i.e., the quiescent fraction as a function of M$_{\rm{star}}$) does change. The top panel of Figure~\ref{fig:norm} shows that the quiescent fraction significantly increases at different M$_{\rm{star}}$ values as a result of the change in the normalization of the M$_{\rm{BH}}$-M$_{\rm{star}}$ relation. This indicates changes to the extent of the star forming main sequence on a sSFR-M$_{\rm{star}}$ plot.

We find, however, that the observational sample from T17 exhibits a larger intrinsic scatter in the M$_{\rm{BH}}$-M$_{\rm{star}}$ relation that is not produced by TNG, as discussed in Section~\ref{sec:results}. We confirm that lowering the seed M$_{\rm{BH}}$ only increases the scatter for low M$_{\rm{star}} \lesssim 10^{10.2}$ M$_{\odot}$, above which the scatter remains small. This discrepancy likely comes from the particular circumstances in which black holes coevolve and grow with their host galaxies in TNG (Li et al. in prep). The broad scatter in the observational sample's M$_{\rm{BH}}$-M$_{\rm{star}}$ relation suggests that black holes can exhibit different mass assembly histories in different galaxies. If M$_{\rm{BH}}$ is truly an indicator of quiescence, then the intrinsic scatter in how galaxies can host black holes of different masses will largely affect the observed relationship between sSFR and M$_{\rm{star}}$.

\subsection{The M$_{\rm{BH}}$ threshold for quiescence}
\label{sec:threshold}

The top middle panel of Figure~\ref{fig:3Dcube} shows that for TNG galaxies with M$_{\rm{star}}  > 10^{10}$ M$_{\odot}$ there exists a M$_{\rm{BH}}$ threshold for quiescence at $\sim10^{8.2}$ M$_{\odot}$ where below this mass most ($>$90\%) of these galaxies are star-forming and above this mass most ($>$90\%) are quiescent. The majority of these quiescent central galaxies (73\%) have very low sSFRs shown as upper limits at $\sim10^{-12.5}$ yr$^{-1}$, demonstrating that black hole kinetic winds are extremely effective at preventing star formation in most galaxies undergoing this form of black hole feedback in TNG. In this section, we show that the M$_{\rm{BH}}$ threshold at $\sim10^{8.2}$ M$_{\odot}$ is sensitive to the parameter choices that constitute the TNG fiducial model.

The four model variations that illustrate this point are the HighMpiv, LowMpiv, HighChi0, LowChi0 models (see Table~\ref{tab:models}). In Section~\ref{sec:bhs}, we describe how TNG uses a M$_{\rm{BH}}$-dependent Eddington ratio threshold (Eqn.~\ref{eqn:chi}) for determining whether a black hole is releasing thermal or kinetic mode feedback energy. Each of the four model variations we analyze in this section changes the Eddington-ratio threshold, $\chi$, by a certain factor, $\eta$:
\begin{equation}
\begin{aligned}
\chi = \rm{min}\bigg[ \eta \chi_{\rm{fid}}, \chi_{\rm{max}}\bigg],
\label{eqn:chi2}
\end{aligned}
\end{equation}
where $\chi_{\rm{fid}}$ is the fiducial M$_{\rm{BH}}$-dependent Eddington ratio threshold (see Eqn.~\ref{eqn:chi}), $\chi_{\rm{max}} = 0.1$ as described in Section~\ref{sec:bhs}, and $\eta = \frac{1}{16}, \frac{1}{4}, 4, 16$ for the HighMpiv, LowChi0, HighChi0, and LowMpiv model variations, respectively. These variations change the normalization of the threshold. 

In Figure~\ref{fig:boundary} we show the Eddington ratio as a function of M$_{\rm{BH}}$ for these four model variations (median distributions shown as solid blue lines), explicitly showing the changes to the Eddington ratio threshold between thermal and kinetic mode feedback (shown as dotted blue lines). The variation in the Eddington ratio threshold changes how galaxies populate this parameter space. 

The changes to the normalization of the Eddington ratio threshold change the M$_{\rm{BH}}$ at which black hole kinetic wind energy can begin to accumulate. Since kinetic winds from black holes are necessary to produce quiescence in TNG (see Section~\ref{sec:quiescence}), changes in the normalization of $\chi$ result in changes to the M$_{\rm{BH}}$ threshold for quiescence since $\chi$ is M$_{\rm{BH}}$-dependent.

This behavior is shown in Figure~\ref{fig:mbhthreshold}, where the M$_{\rm{BH}}$-M$_{\rm{star}}$ relation is shown for each model variation at $z = 0$. From left to right, the M$_{\rm{BH}}$ threshold at which galaxies become mostly quiescent moves down as the normalization of $\chi$ increases. The quiescent fraction as a function of M$_{\rm{star}}$ and M$_{\rm{BH}}$ are shown on the top and right panels of each plot, respectively. As the M$_{\rm{BH}}$ threshold for quiescence decreases, the M$_{\rm{star}}$ at which galaxies begin exhibiting quiescence also decreases, until $\chi = 16\chi_{\rm{fid}}$ (rightmost panel) where most of the population above M$_{\rm{star}} = 10^{10}$ M$_{\odot}$ is quiescent and above this threshold.

We also note that the number of galaxies at high M$_{\rm{BH}}$ and M$_{\rm{star}}$ decreases as $\chi$ increases from left to right in Figure~\ref{fig:mbhthreshold}. This is because increasing the $\chi$ value makes it easier for black holes to satisfy the conditions necessary for kinetic mode feedback to be turned on. Namely, black holes do not need to be as massive or accrete as inefficiently as would be necessary with a lower $\chi$ value. This increases the amount of time that galaxies spend in the kinetic mode since lower mass systems can become quiescent earlier. We have shown in Section~\ref{sec:quiescence} that kinetic winds are necessary to produce quiescence and therefore largely stop the growth of galaxies via \textit{in situ} star formation. Concurrently, kinetic winds also largely halt the growth of the black hole as can be seen in Figure~\ref{fig:boundary}, where the Eddington ratio decreases dramatically once galaxies enter the kinetic mode \citep{wsp2018, hgs2019}. Therefore, massive black holes and massive galaxies become rarer in the model variations with larger $\chi$ values due to the inefficient growth of M$_{\rm{BH}}$ and M$_{\rm{star}}$ for these galaxies.

We note that the M$_{\rm{BH}}$ threshold for quiescence at $\sim10^{8.2}$ M$_{\odot}$ emerges from the ensemble of parameter choices that produce a realistic $z = 0$ galaxy population. TNG model parameters were calibrated in order approximately return the shape and amplitudes of certain observables mostly related to the stellar mass content of $z = 0$ galaxies \citep[see][]{psn2018}. While the model variations shown in Figure~\ref{fig:mbhthreshold} produce unrealistic galaxy populations that would violate the observational constrains described in \citet{psn2018}, they demonstrate the sensitivity of galaxy properties to changes in the M$_{\rm{BH}}$ threshold.

While different TNG model variations exhibit different behaviors in the sSFR-M$_{\rm{BH}}$-M$_{\rm{star}}$ parameter space, we note that the binding energy correlation described in Section~\ref{sec:phenom} continues to hold. Namely, for all model variations, the accumulated energy from black hole-driven kinetic winds must exceed the binding energy of gas within the galaxy in order for quiescence to be produced in each model (See Appendix~\ref{sec:physicsmvs}). Differences in the model variations occur when the galaxy properties required to meet this condition are changed. This occurs, for example, by allowing lower mass black holes to accumulate kinetic feedback energy (as in this section), or by changing the way black holes occupy galaxies of different M$_{\rm{star}}$ (as in Section~\ref{sec:coevol}).

\section{Discussion}
\label{sec:disc}

This work aims to illuminate how the effects of black hole feedback on gas within and around galaxies can alter the observable properties of sSFR, M$_{\rm{BH}}$, and M$_{\rm{star}}$ in the context of a physical model of galaxy formation. We have used the IllustrisTNG simulation suite and its model variations to characterize and assess the black hole model and the resulting physics of quiescence for central galaxies with M$_{\rm{star}} > 10^{10}$ M$_{\odot}$. In this section, we discuss how our results may aid in understanding the link between black hole feedback and quiescence in the real Universe, and how other approaches to modeling this feedback in current and future simulations may benefit from the type of analysis we do in this work.

\subsection{Assessing models using the sSFR-M$_{\rm{BH}}$-M$_{\rm{star}}$ parameter space}

We turn our attention first to the sharp transition from a mostly star-forming to a mostly quiescent galaxy population above a particular M$_{\rm{BH}}$ threshold at $\sim10^{8.2}$ M$_{\odot}$ in TNG. We compare this to a sample of 91 galaxies with dynamical M$_{\rm{BH}}$ measurements from T17. We note that this sample is not representative of the entire galaxy population due to the biases associated with various dynamical M$_{\rm{BH}}$ measurement techniques.

The current data available, however, do not indicate the existence of a particular M$_{\rm{BH}}$ threshold for quiescence as is seen in TNG (see the middle panel of Figure~\ref{fig:3Dcube}). Instead, these data show that the sSFR is a smoothly decreasing function of the M$_{\rm{BH}}$/M$_{\rm{star}}$ ratio of these galaxies \citep[for a full discussion see][]{tbw2017}. This empirical relationship provides a constraint for models that use black hole feedback in order to produce quiescence. Future work using more representative samples of dynamical M$_{\rm{BH}}$ measurements with an understanding of their selection biases will be important for more precisely quantifying the observed correlations between sSFR, M$_{\rm{star}}$, and M$_{\rm{BH}}$ found in T17.

Another point of discussion centres on the fact that TNG produces a M$_{\rm{BH}}$-M$_{\rm{star}}$ relation with significantly less intrinsic scatter compared to the observational data from T17 (See the left panel of Figure~\ref{fig:3Dcube}). Earlier studies of black hole demographics were heavily biased towards the most massive black holes and mostly took into account those within quiescent, early-type galaxies and a few bulge components of late-type galaxies \citep{mtr1998, fm2000, gbb2000}. The tight correlation between black holes and the galaxy properties of their early-type components led to the use of samples like these \citep[e.g., Figure 2 of][ showing the M$_{\rm{BH}}$-M$_{\rm{bulge}}$ relation for early-type galaxies]{mm2013} to assess a model's black hole demographics for the entirety of the galaxy population, early-types or otherwise. The small scatter seen in TNG's M$_{\rm{BH}}$-M$_{\rm{star}}$ relation is a common feature in other galaxy formation models for similar reasons \citep[e.g., Section 3.3 of][]{vdp2016}.

While recent studies of black hole demographics using M$_{\rm{BH}}$ measurements are still biased due to the specifics of each detection method, they incorporate a more diverse set of galaxies and show a significantly more pronounced intrinsic scatter in the relationship between M$_{\rm{BH}}$ and M$_{\rm{star}}$ \citep{rv2015, sgm2015, tbh2016b, tbw2017, dgc2018, sgd2019}. Namely, they show that late-type, star-forming galaxies lie below the canonical M$_{\rm{BH}}$-M$_{\rm{star}}$ relations for early-type, quiescent galaxies, increasing the scatter in the relation for the whole population (see also Li et al. in prep). Models need to take both the large scatter and its correlation with star formation properties into account when assessing whether their black hole demographics adequately represent the whole galaxy population.

The coevolution of M$_{\rm{star}}$ and M$_{\rm{BH}}$ is difficult to constrain, especially since the processes governing the growth of these two properties function at spatial and temporal scales that differ by several orders of magnitude. Observations show that M$_{\rm{BH}}$ accretion rates differ for galaxies of different masses and star formation properties \citep{acm2012, acg2017, acg2018}. Additionally, studies have shown that different black hole seeding mechanisms can affect the evolution of both the black hole and its host galaxy \citep{wtf2019}. These factors are likely important for determining the intrinsic scatter between M$_{\rm{star}}$ and M$_{\rm{BH}}$.

Uncertainties on how to model accretion onto a black hole and how this relates to large-scale gas accretion onto its host galaxy also exacerbates the issue of linking M$_{\rm{star}}$ and M$_{\rm{BH}}$ in simulations. Many studies argue that Bondi-Hoyle accretion \citep{hl1939, bh1944} is an oversimplification, assuming the gas immediately surrounding the black hole has zero angular momentum. This assumption fails to take into account the fact that non-spherical accretion is an important factor for growing black holes \citep[e.g.,][]{hq2011}. While Bondi-Hoyle accretion is an attractive model due to its simplicity, it may prove to be an important limitation when attempting to reproduce large-scale galaxy properties. Other models for black hole growth, such as the gravitational torque model \citep{hq2011, aod2015, adf2017, afq2017}, will provide different avenues for exploring possible black hole-galaxy coevolutionary scenarios once they are fully incorporated into a large-scale cosmological model of galaxy formation that includes black hole feedback.

Black hole feedback itself can also affect the growth of black holes in galaxies. In Section~\ref{sec:coevol}, we show that the thermal mode feedback implemented at high accretion rates in TNG regulates M$_{\rm{BH}}$ growth and plays a role in setting the normalization of the M$_{\rm{BH}}$-M$_{\rm{star}}$ relation. Figure~\ref{fig:boundary} also shows that the Eddington ratio drops to low values once the distribution of galaxies on this plot crosses the threshold where kinetic mode feedback takes effect \citep{wsp2018}. \citet{hgs2019} explore the black hole population and the the properties of AGN in TNG, showing that their kinetic wind feedback implementation may be too effective at suppressing M$_{\rm{BH}}$ growth compared to observational constraints. Additionally, TNG assumes constant coupling ($\varepsilon_{\rm{f,high}}$, $\varepsilon_{\rm{f,kinetic}}$) and radiative ($\varepsilon_{\rm{r}}$) efficiencies for black hole feedback (see Section~\ref{sec:bhs} and \citealt{wsp2018}). The values for these efficiencies likely vary across the galaxy population and would increase the scatter in the M$_{\rm{BH}}$-M$_{\rm{star}}$ relation \citep{bs2019}.

Stellar feedback can also slow M$_{\rm{BH}}$ growth by reducing the amount of cold gas present near the black hole, particularly in shallow potential wells during the early growth of the black hole \citep{hvd2017, mbh2017, psn2018}. At the galaxy mass scales relevant to this work with M$_{\rm{star}} > 10^{10}$ M$_{\odot}$, stellar feedback is also expected to play an important role in setting the mass scale above which quiescent galaxies begin to dominate the galaxy population. \citet{bsf2017} and \citet{hwl2019} both demonstrate this by showing that black hole activity can be triggered once stellar feedback becomes ineffective at preventing cooling from the gaseous halo to the galaxy in progressively more massive systems.

The prescriptions for black hole accretion, the implementation of stellar and black hole feedback, and the complex interactions between feedback's effect on the availability of gas for M$_{\rm{BH}}$ and M$_{\rm{star}}$ growth all contribute to the characteristics of simulated M$_{\rm{BH}}$-M$_{\rm{star}}$ relations. In this work, we show that if M$_{\rm{BH}}$ is causally related to quiescence, then the shape, normalization, and scatter of the M$_{\rm{BH}}$-M$_{\rm{star}}$ relation will determine the M$_{\rm{star}}$ distributions of star-forming and quiescent galaxies (e.g., in TNG compared to Illustris, \citealt{dpn2019}, Li et al. in prep). This is a fundamental observational diagnostic for models. As such, it is essential that the the observed scatter in the M$_{\rm{BH}}$-M$_{\rm{star}}$ relation and its correlation with its host galaxy's sSFR are reproduced for any model where M$_{\rm{BH}}$ correlates with quiescence. This will largely determine the stellar mass distribution of the star forming main sequence and the galaxies that lie off of it, as we describe in Section~\ref{sec:implications}.

\subsection{Characterizing the impact of black hole kinetic winds in TNG}

In Section~\ref{sec:phenom}, we show that quiescent galaxies have black hole wind energies that exceed the binding energy of the gas in the galaxy. This supports a framework where black hole kinetic winds gravitationally unbind gas from the galaxy above a M$_{\rm{BH}}$ threshold of $\sim10^{8.2}$ M$_{\odot}$.

Following our analysis in Section~\ref{sec:phenom} and visually indicated by Figure~\ref{fig:gasdistcomp}, black hole winds push dense gas out of galaxies in TNG, effectively producing quiescence. However, high density, cooler gas is more difficult to gravitationally unbind or heat with feedback energy than gas that is more diffuse. Other studies confirm that energy from these winds in TNG galaxies affects not only the cold gas but also the more diffuse gas that preferentially resides at high scale heights above the galaxy and within the circumgalactic medium \citep[][Zinger et al. in prep]{nkp2018, knb2019}.

This likely explains the behavior seen in the bottom panel of Figure~\ref{fig:modelcomp}, where galaxies just above the M$_{\rm{BH}}$ threshold for quiescence exhibit an abrupt decrease in halo gas mass. More massive black holes that live in more massive haloes are able to retain a larger fraction of their gas since it is increasingly difficult to push gas out of a deeper potential well. \citet{dcm2019} and \citet{odc2019} independently found similar behavior in the EAGLE simulation, where the gas fraction decreases when the ratio of black hole feedback energy and halo binding energy is high.

Figures 4 and 8 in \citet{psn2018} and Figure 11 in \citet{wsh2017} show rough agreement with observational constraints of halo gas masses around galaxies with dark matter halo masses $>10^{13}$ M$_{\odot}$, showing an improvement compared to the original Illustris model where halo gas masses were severely underestimated. However, while there is evidence to suggest that halo gas masses must be suppressed by feedback in order to agree with cluster mass estimates \citep[e.g.,][]{con2015, con2017, bkb2017}, the evacuation of $\gtrsim80$\% of halo gas for a large population of intermediate mass galaxies undergoing black hole feedback should be corroborated before the agreement between TNG and observations is taken at face value. Upcoming work will compare the the observational X-ray signatures of the gaseous atmospheres in TNG haloes with observational data (Pop et al. in prep, Truong et al. in prep).

Within the framework of ejective and preventative feedback, TNG's kinetic wind model produces quiescence through both mechanisms. The idea that TNG produces quiescence through sustained ejective feedback is supported by our analysis in Section~\ref{sec:sam} where we find that quiescence in TNG cannot be easily described using the logic of balancing heating and cooling rates employed by purely preventative semi-analytic models. However, Figure~\ref{fig:modelcomp} shows that the cooling time in galaxies exhibiting black hole winds is significantly greater than in the model variation where no black hole winds are present. \citet{wsh2017} demonstrate that these winds are able to thermalize at large distances from the black hole, indicating that the kinetic mode black hole feedback mechanism employed in TNG is partly preventative, resulting in the increase of the gas halo's average cooling time. This is further supported by results from \citet{nps2019} who show the thermal properties of gas around a massive galaxy in TNG50 undergoing kinetic black hole feedback (see their Figure 2 and Zinger et al. in prep). Additionally, the ejective nature of TNG's black hole kinetic winds likely aids the process that leads to quiescence since it reduces the mass of the reservoir from which star-forming gas is accreted onto the galaxy.

\subsection{Black hole feedback prescriptions in models}

Another key issue for large-volume simulations is the need to implement subgrid physics that models the small-scale phenomena important for galaxy evolution. One crucial approximation is how energy from feedback is transferred to the surrounding gas. Feedback from both stars and black holes has been modeled using a variety of different subgrid physics. The details of these subgrid physics will inevitably affect observable galaxy properties and the transfer of feedback energy to the gas within and around galaxies.

In the case of TNG, we have described a two-mode black hole feedback model separating the forms of energy injection at high- and low-accretion rates, corresponding to radiatively efficient and inefficient accretion. High- and low- accretion phases have been modeled in TNG as an injection of pure thermal energy or pure kinetic energy, respectively. The low-accretion phase injection of kinetic winds in TNG was motivated by recent observational evidence of strong outflows in galaxies hosting inefficiently accreting black holes \citep{cbc2016, wsz2017, pms2018}. Theoretically, this implementation was also motivated by the wind launching mechanism from inefficient accretion onto black holes put forth by studies such as \citet{bb1999}.

In this work, we show that thermal mode black hole feedback in TNG regulates M$_{\rm{BH}}$ growth (Section~\ref{sec:coevol}) whereas kinetic mode feedback suppresses M$_{\rm{BH}}$ and, most notably for this work, M$_{\rm{star}}$ growth (Sections~\ref{sec:quiescence} and~\ref{sec:physics}). The abrupt change in galaxy and halo gas properties at the particular M$_{\rm{BH}}$ threshold we describe in this work (e.g., Figure~\ref{fig:modelcomp}) indicates an abrupt change in physical processes affecting galaxies that are undergoing kinetic mode as opposed to thermal mode feedback. Similar two-mode black hole feedback models have been implemented in a number of other large-volume simulations \citep[e.g.][]{csw2006, shc2008, vgs2014, hwt2015, dpp2016}.

However, this dichotomy between feedback modes is likely an oversimplification of the physics that occurs in the real Universe. As discussed in Section~\ref{sec:intro}, there is an abundance of observational evidence for supermassive black hole activity in galaxies. However, the interpretation of this evidence for constructing a generalizable black hole feedback model for all forms of accretion is not straightforward. For example, radio jets and lobes, generally attributed to low accretion rate feedback, have also been seen in galaxies that are producing large-scale outflows from supposedly high accretion rates \citep[e.g.,][]{kvx2006, bcj2018}.

While TNG produces winds through low rates of accretion, some higher resolution zoom-in simulations that explicitly include kinetic feedback from high rates of accretion have shown that this process can also effectively eject cold gas from galaxies, depress the central density of the circumgalactic medium, and reduce star formation in galaxies over long timescales \citep[e.g.,][]{con2012, con2015, con2017, cso2018}. This is supported by observational evidence of outflows in AGN and quasars \citep[e.g.,][]{hmv1981, vhd2011, cms2014}. These types of zoom-in studies take advantage of their smaller volumes in order to incorporate multiple avenues of feedback energy transfer from black holes (e.g., kinetic, thermal, and/or radiative) simultaneously that may more comprehensively model the complex interactions between black holes and the ambient medium \citep[e.g.,][]{bs2017, mbw2018, bcs2018}. 

As such, a consensus on how black hole feedback should be implemented in cosmological simulations within the limitations of finite resolution has not yet been reached. These complications need to be considered when assessing models that simplify black hole feedback physics into low- and high-accretion rate modes. This dichotomy is likely missing important physical recipes that may shape the way galaxies grow in simulations.

\section{Conclusions}
\label{sec:conclusions}

We explore the effects of black hole feedback on the properties of central galaxies with M$_{\rm{star}}  > 10^{10}$ M$_{\odot}$ in the context of the IllustrisTNG simulation suite. In particular, we use TNG100 and ten model variations to assess how observable correlations between the sSFR, M$_{\rm{star}}$, and M$_{\rm{BH}}$ of these galaxies are sensitive to changes in the physics model. We also connect these correlations to the effects of black hole feedback on the distribution of gas within and around galaxies. Finally, we compare results from TNG with observational data of galaxies with dynamical M$_{\rm{BH}}$ measurements. We highlight our main results below:

\begin{itemize}

\item TNG requires low accretion rate black hole feedback in the form of kinetic winds in order to produce a quiescent galaxy population (Section~\ref{sec:quiescence}, Figure~\ref{fig:ssfrhist}, and also shown in \citealt{wsh2017}).

\item A decline in the sSFR of simulated galaxies is seen when the accumulated black hole wind energies exceed the gravitational binding energies of the gas within galaxies, $\int \dot{E}_{\rm{kinetic}} dt > E_{\rm{bind, gal}}$ (top panel of Figure~\ref{fig:energyrates}). This behavior is seen for all model variations we examine in this work (See Appendix~\ref{sec:physicsmvs}). This provides strong evidence that black hole-driven kinetic winds push cold gas out of the galaxy to produce quiescence in TNG (Section~\ref{sec:phenom}, Figure~\ref{fig:gasdistcomp}). 

\item Simulated galaxies with black hole wind energies that fall below the binding energy of gas within the galaxy have black hole masses below a M$_{\rm{BH}}$ threshold of $\sim10^{8.2}$ M$_{\odot}$. Those that have wind energies exceeding the binding energies host black holes above this M$_{\rm{BH}}$ threshold (bottom panel of Figure~\ref{fig:energyrates}). This produces a sharp decrease in the amounts of interstellar and circumgalactic gas at this M$_{\rm{BH}}$ threshold (Section~\ref{sec:icgm}, Figure~\ref{fig:modelcomp}).

\item Below the M$_{\rm{BH}}$ threshold at $\sim10^{8.2}$ M$_{\odot}$, most ($>$90\%) simulated central galaxies with M$_{\rm{star}}  > 10^{10}$ M$_{\odot}$ are star-forming and above this mass most ($>$90\%) are quiescent. 73\% of the quiescent population have very low sSFRs (with upper limits at $10^{-12.5}$ yr$^{-1}$), indicating that black hole winds are extremely effective at suppressing star formation once low-accretion rate feedback takes effect in TNG (also see \citealt{wsp2018}). 

\item We compare TNG to observational data of 91 central galaxies with dynamical M$_{\rm{BH}}$ measurements from \citet{tbh2016b, tbw2017}, with the caveat that these data are not representative of the entire galaxy population. We find that TNG qualitatively reproduces the result that quiescent galaxies host more massive black holes than star-forming galaxies. However, the M$_{\rm{BH}}$-M$_{\rm{star}}$ relation for TNG produces a much smaller scatter compared to what is seen in the observational data (Section~\ref{sec:results}, top left panel of Figure~\ref{fig:3Dcube}). Additionally, these observational data, while incomplete, show a smoother decline in sSFR as a function of M$_{\rm{BH}}$, showing no indication of an abrupt suppression of sSFR at a particular M$_{\rm{BH}}$ threshold (Section~\ref{sec:results}, top middle and bottom panels of Figure~\ref{fig:3Dcube}).

\item The distribution of star-forming and quiescent galaxies across M$_{\rm{star}}$ parameter space in TNG is sensitive to both the normalization of the M$_{\rm{BH}}$-M$_{\rm{star}}$ relation and the M$_{\rm{BH}}$ at which black holes produce kinetic winds. We show that the normalization of the M$_{\rm{BH}}$-M$_{\rm{star}}$ relation depends on the efficiency of thermal mode feedback on regulating M$_{\rm{BH}}$ growth.

\end{itemize}

These results demonstrate that the relationship between M$_{\rm{BH}}$, M$_{\rm{star}}$, and sSFR is a powerful tool for exploring the physics of quiescence within the context of black hole feedback. We show that if the M$_{\rm{BH}}$ and sSFR of galaxies are causally linked, as is the case in TNG, then the way M$_{\rm{BH}}$ pairs with galaxies of different M$_{\rm{star}}$ will determine the star formation properties of the entire central galaxy population. We find important differences between the results from TNG and the observational data for galaxies with dynamical M$_{\rm{BH}}$ measurements from \citet{tbh2016b, tbw2017}. These differences illuminate the importance of taking into account the scattered relationship between M$_{\rm{BH}}$ and M$_{\rm{star}}$ and its dependence on sSFR found in current samples of galaxies with dynamical M$_{\rm{BH}}$ measurements.

\section*{Acknowledgements}

B.A.T. is supported by the National Science Foundation Graduate Research Fellowship under Grant No. DGE1256260. The Flatiron Institute is supported by the Simons Foundation. The flagship simulations of the IllustrisTNG project used in this work have been run on the HazelHen Cray XC40-system at the High Performance Computing Center Stuttgart as part of project GCS-ILLU of the Gauss centres for Supercomputing (GCS). Ancillary and test runs of the project were also run on the Stampede supercomputer at TACC/XSEDE (allocation AST140063), at the Hydra and Draco supercomputers at the Max Planck Computing and Data Facility, and on the MIT/Harvard computing facilities supported by FAS and MIT MKI. In particular, the model variation runs adopted in this study have been run on the Draco supercomputer. We thank David Barnes, Rahul Kannan, Kareem El-Badry, Mariska Kriek, Kayhan G{\"u}ltekin, August Evrard, Elena Gallo, Brian O'Shea, and Chung-Pei Ma for helpful discussions.



\bibliographystyle{mnras}
\bibliography{paper}



\appendix

\section{The Physics of Quiescence in the Model Variations}
\label{sec:physicsmvs}

%
\begin{figure*}
\centering
\epsfxsize=16.5cm
\epsfbox{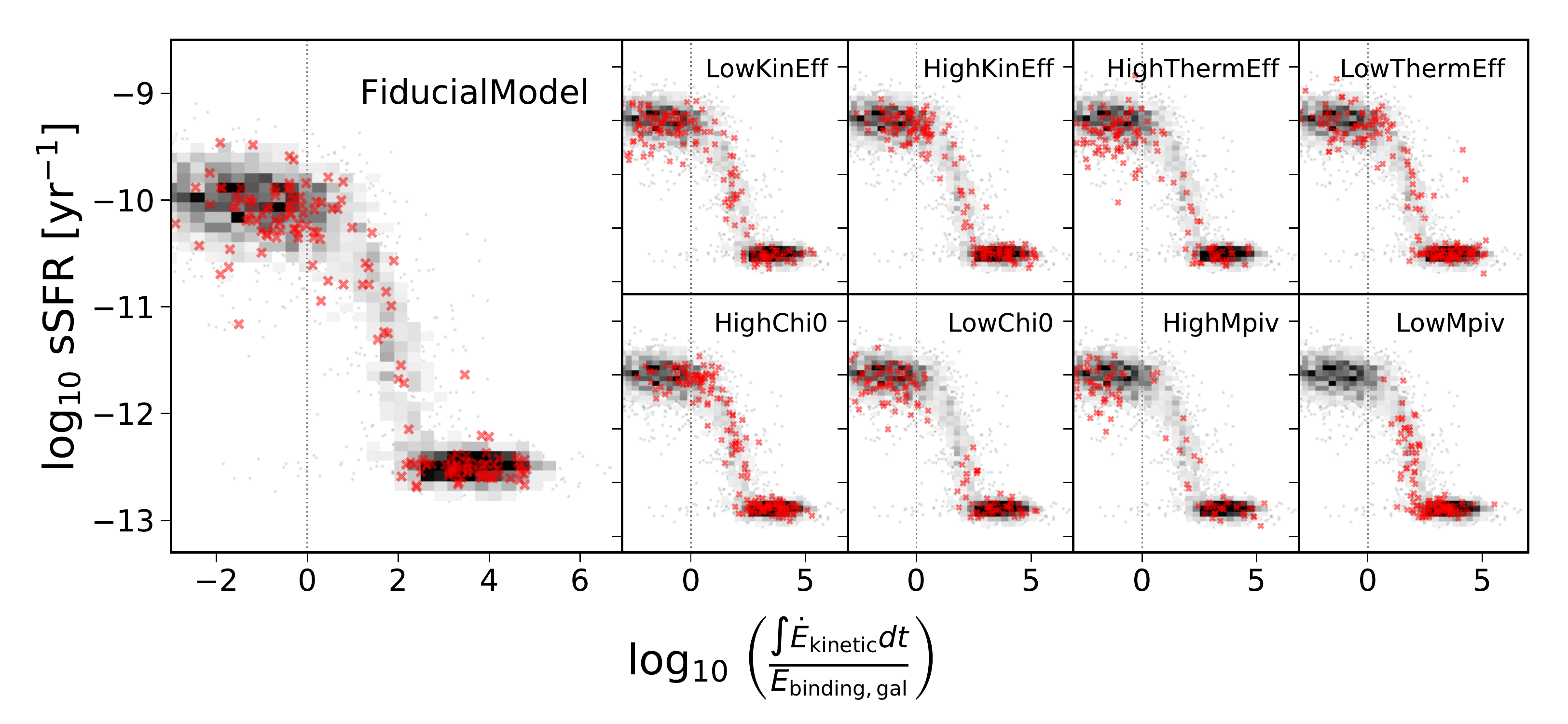}
\caption{The sSFR as a function of the ratio between the cumulative energy from black hole-driven kinetic winds and the binding energy of gas in the galaxy for all model variations listed in Table~\ref{tab:models} (red crosses). The gray heatmaps are the same in each panel and show the distribution of galaxies in TNG100. All models reproduce the same relation as the fiducial model on this plot.}
\label{fig:allmvs}
\end{figure*}

Here we append further evidence in support of our phenomenological framework presented in Section~\ref{sec:phenom}. This framework describes a physically-motivated picture of quiescence in TNG, where a majority of galaxies shut off their star formation once black hole-driven winds accumulate enough energy to push gas out of their host galaxy. This framework comes from the fact that the sSFR of a galaxy declines once the cumulative energy from black hole-driven winds becomes larger than the binding energy of gas in the galaxy.

The model variations listed in Table~\ref{tab:models} alter the parameters of the black hole physics model in TNG, producing significantly different galaxy populations. This is described in Section~\ref{sec:implications} and shown in Figures~\ref{fig:boundary},~\ref{fig:norm}, and~\ref{fig:mbhthreshold}. Even so, the evidence for our phenomenological framework holds true for every model variation we assess in this work. In Figure~\ref{fig:allmvs}, we show the sSFR as a function of the ratio between accumulated black hole wind energy and the binding energy of gas in the galaxy (the same parameter space as the top panel of Figure~\ref{fig:energies}). Each panel shows the distribution of TNG100 galaxies as a grayscale heatmap, while the individual galaxies in each model variation are shown as red crosses. The fiducial TNG model is shown in the large panel on the left, whereas the eight model variations that include black hole-driven kinetic winds are to the right.

Despite the differences in galaxy population statistics in each model variation, we find that galaxies in all of the simulations reproduce the same relation on these plots as the fiducial model runs. The only differences are where galaxies lie along the distribution compared to one another. For example, in Section~\ref{sec:threshold} and Figure~\ref{fig:mbhthreshold} we show that most of the galaxies in the HighMpiv model variation are star-forming since the criteria for producing kinetic winds is only satisfied by the most massive black holes and therefore the most massive galaxies. For the LowMpiv model variation, the criteria for producing kinetic winds is satisfied by all galaxies with M$_{\rm{star}} > 10^{10}$ M$_{\odot}$. In Figure~\ref{fig:allmvs}, the distribution of galaxies in these two model variations reflect this behavior since in the LowMpiv model all black holes produce enough accumulated energy to overcome the binding energy of the gas within the galaxy. Namely, all galaxies in this simulation lie to the right of the vertical dotted line where these two energies equal. The HighMpiv model shows galaxies that have not yet produced enough black hole wind energy to do this and as such also populate the star-forming region of this plot to the left of the vertical dotted line.

The NoBHs and NoBHwinds simulations do not have a value for $\int \dot{E}_{\rm{kinetic}} dt$. However, we have shown in Figure~\ref{fig:ssfrhist} that they do not produce a population of quiescent galaxies, further supporting our phenomenological framework.

\section{Resolution effects in TNG300 compared with TNG100}
\label{sec:tng300}

In this work, we exclusively use results from TNG100 in order to directly compare with the model variations that have roughly the same resolution. Here we describe resolution effects present in TNG300 that further motivate the exclusive use of TNG100 for the purposes of our study.

TNG300 is one of the flagship simulations of the IllustrisTNG project with a 302.6$^{3}$ comoving Mpc$^{3}$ volume. TNG300 has $2\times2500^{3}$ initial resolution elements with a baryonic mass resolution of $1.1\times10^{7}$ M$_{\odot}$ and gravitational softening length of 1.48 kpc at $z = 0$. This represents approximately an order of magnitude decrease in mass resolution and an increase in the gravitational softening length by a factor of 2 compared to TNG100.

\begin{figure}
\centering
\epsfxsize=7.3cm
\epsfbox{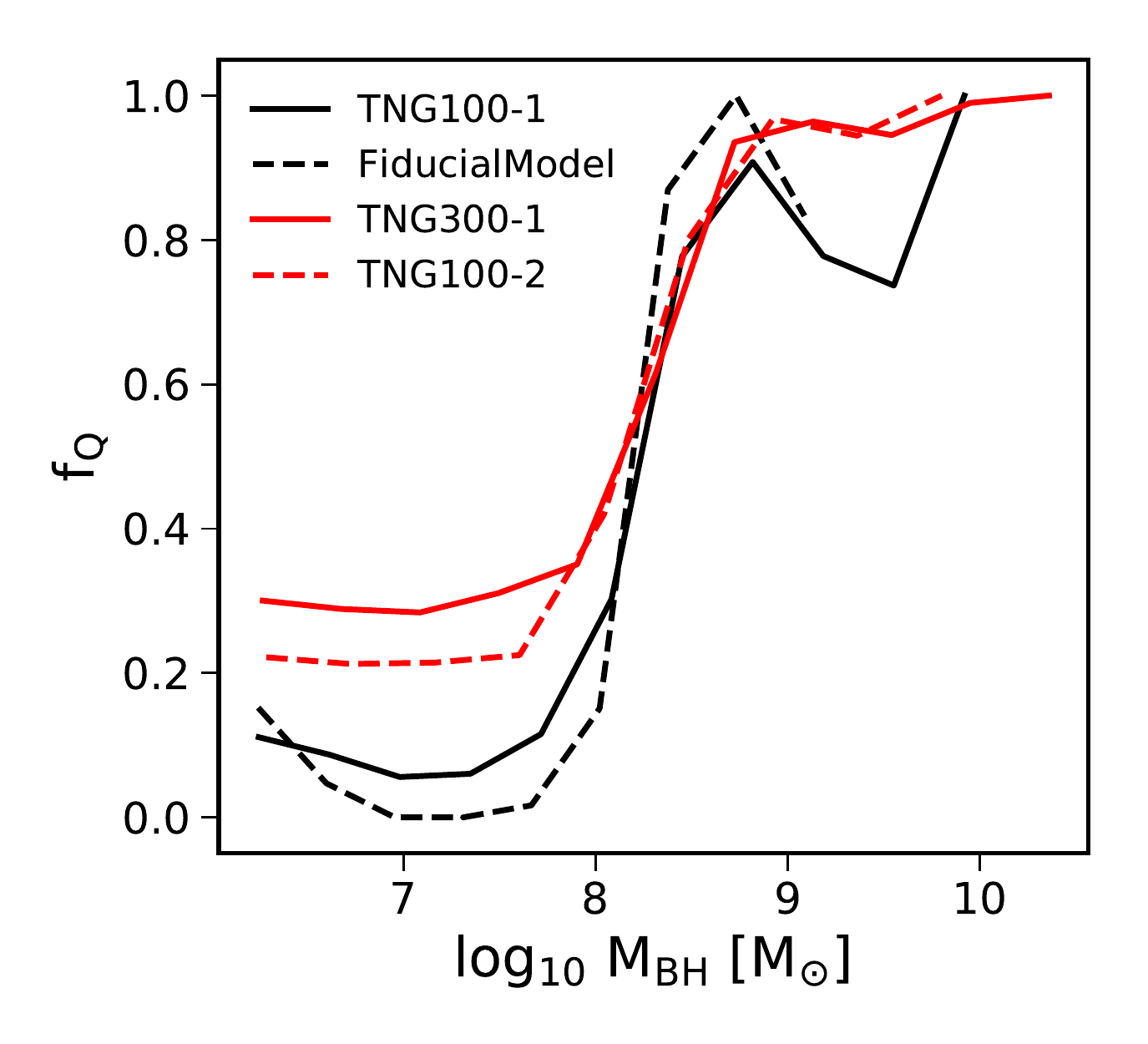}
\caption{The fraction of quiescent galaxies as a function of M$_{\rm{BH}}$ in TNG100-1 (black, solid line), FiducialModel (black, dashed line), TNG300-1 (red, solid line), and TNG100-2 (red, dashed line). The black lines represent simulations run at higher resolution than those represented by the red lines. The dashed lines represent smaller volume simulations than those represented by the solid lines. Lower resolution TNG runs produce up to 20\% more quiescent galaxies at M$_{\rm{BH}} \lesssim 10^{8.2}$ and $\gtrsim 10^{9}$ M$_{\odot}$.}
\label{fig:fqres}
\end{figure}

\begin{figure}
\centering
\epsfxsize=7.3cm
\epsfbox{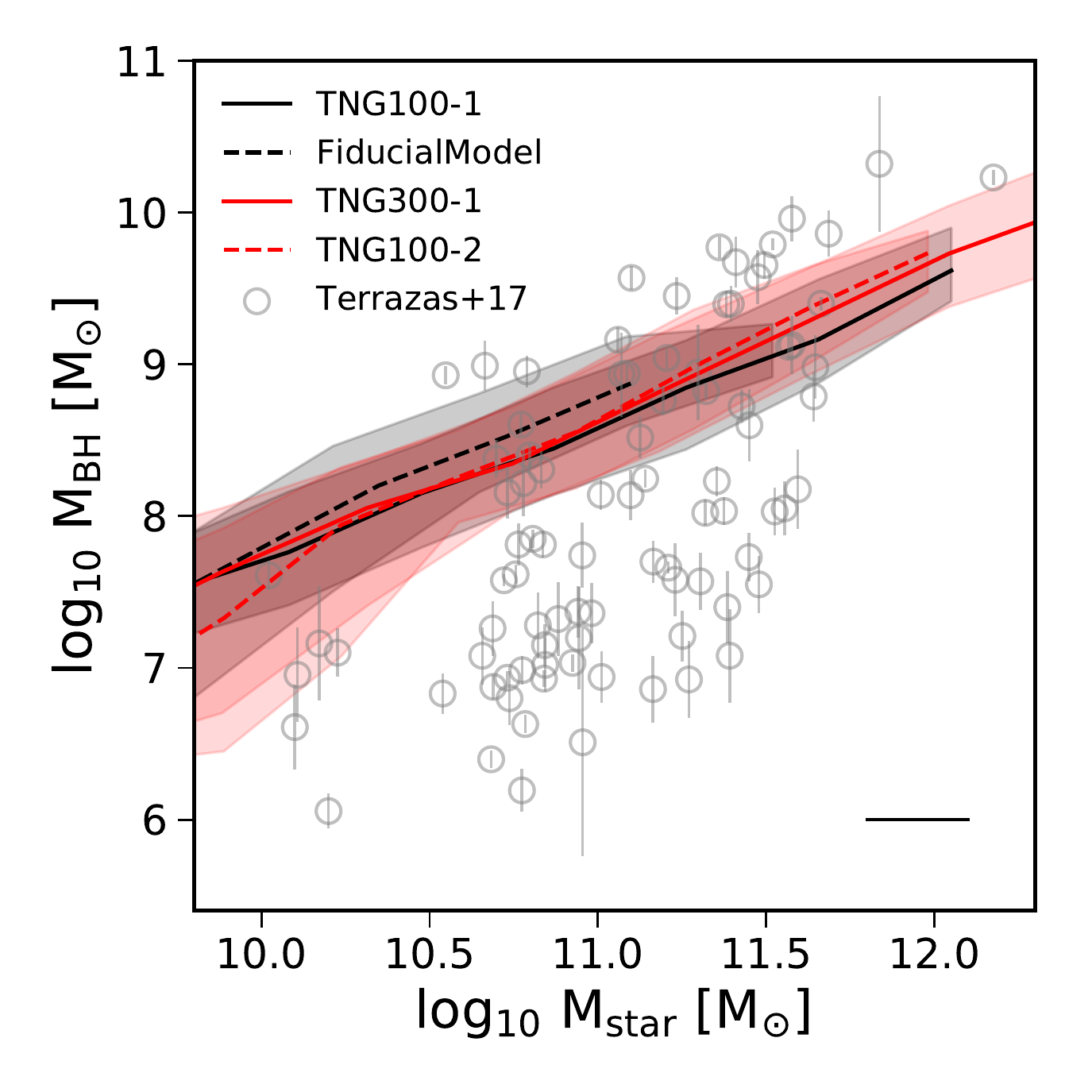}
\caption{The median M$_{\rm{BH}}$-M$_{\rm{star}}$ relation for TNG100-1 (black, solid line), FiducialModel (black, dashed line), TNG300-1 (red, solid line), and TNG100-2 (red, dashed line). The 10 and 90 percentiles below and above the median are shown as translucent regions matching the colour of their median lines. Similar to Figure~\ref{fig:fqres}, black and gray colours represent simulations run at higher resolution than those represented by red and the dashed lines represent smaller volume simulations than those represented by the solid lines. Lower resolution TNG runs produce a slightly more scattered relation at M$_{\rm{star}} \lesssim 10^{10.4}$ M$_{\odot}$. Above this mass, the scatter is consistent between simulations. The observational sample of galaxies with dynamical M$_{\rm{BH}}$ measurements are shown as gray circles and its M$_{\rm{star}}$ error bar is shown in the lower right corner.}
\label{fig:relres}
\end{figure}

In order to differentiate between the effects of a smaller/larger box size in comparison to higher/lower resolution, we use four TNG runs: two with the same resolution as the fiducial TNG100 but different box volumes (TNG100-1 sized at 110.7$^{3}$ Mpc$^{3}$ and FiducialModel sized at 36.9$^{3}$ Mpc$^{3}$) and two with the same resolution as TNG300 but different box volumes (TNG300-1 sized at 302.6$^{3}$ Mpc$^{3}$ and TNG100-2 sized at 110.7$^{3}$ Mpc$^{3}$). In Figures~\ref{fig:fqres} and~\ref{fig:relres} the black and red lines indicate the simulations with higher and lower resolution, respectively. Larger and smaller box sizes are indicated with solid and dashed lines, respectively. 

In Section~\ref{sec:physics}, we describe how quiescence occurs above a particular M$_{\rm{BH}}$ threshold for TNG galaxies. Figure~\ref{fig:fqres} shows the fraction of quiescent galaxies as a function of M$_{\rm{BH}}$ for these four simulations. Regardless of box size, those simulations with lower resolution produce up to 20\% more quiescent galaxies at M$_{\rm{BH}} \lesssim 10^{8.2}$ and $\gtrsim 10^{9}$ M$_{\odot}$. This indicates that the feedback present in the lower resolution runs is more effective at suppressing star formation and producing quiescence despite the fact that they have the same subgrid prescriptions. Additionally, star formation itself may also be less efficient at lower resolution.

Our analysis in Section~\ref{sec:results} indicates that the scatter in the M$_{\rm{BH}}$-M$_{\rm{star}}$ relation is much smaller than the scatter seen in the observations. To determine whether the scatter would increase for a larger box simulation, we show the median M$_{\rm{BH}}$-M$_{\rm{star}}$ relations and their 10 and 90 percentile distributions around their medians for the four simulations we discuss in this section. We convolve these M$_{\rm{star}}$ and M$_{\rm{BH}}$ values with observational errors just as we do in Section~\ref{sec:compobs}.

Regardless of box size, those simulations with lower resolution produce a slightly larger scatter than the higher resolution simulations at M$_{\rm{star}} \lesssim 10^{10.4}$ M$_{\odot}$, indicating that this increase in scatter is a resolution effect. Above this mass, the scatter is similarly small for all simulations. We also show the observational data for 91 galaxies with dynamical M$_{\rm{BH}}$ measurements to show that in all these models, the scatter does not easily reproduce the broad distribution of galaxies in the \citet{tbw2017} sample.


\bsp	
\label{lastpage}
\end{document}